\documentclass[
 amsmath,amssymb,
 aps,
]{revtex4-2}

\usepackage{graphicx}
\usepackage{dcolumn}
\usepackage{bm}
\graphicspath{{figures/}}
\usepackage{caption}
\usepackage{subcaption}
\usepackage{ulem}
\usepackage{graphicx}
\usepackage{subcaption}
\usepackage{xcolor}

\usepackage[pagewise]{lineno}
\begin{document}
\preprint{APS/123-QED}
\title{Self-Sustained Oscillations in a Low Viscosity Round Jet}

\author{V. Srinivasan}
\email{vinods@umn.edu}
\author{X. Tan}
\author{E. Whitely}
\author{I. Wright}
\author{A. Dhotre}
\author{J. Yang}
\affiliation{Department of Mechanical Engineering, University of Minnesota, Minneapolis, MN 55455, USA}
\date{\today}

\begin{abstract}
The effect of viscosity contrast between a jet and its surroundings is experimentally investigated using density-matched fluids. A gravity-driven flow is established with a jet of  saltwater emerging into an ambient medium composed of high-viscosity propylene glycol. Jet Reynolds numbers, $Re$, ranging from 1600 to 3400 were studied for an ambient-to-jet viscosity ratio, $M$, between 1 and 50. Visualization suggests that at low values of the viscosity ratio, the jet breakdown mode is axisymmetric, while helical modes develop at high values of viscosity ratio. The transition between these two modes is attempted to be delineated using a variety of diagnostic tools. Hot film anemometry measurements indicate that the onset of the helical mode is accompanied by the appearance of a discrete peak in the frequency spectrum of velocity fluctuations, which exhibits little spatial variation for the first several diameters in the downstream direction. Laser-Induced Fluorescence (LIF) is used to identify the jet boundary against the background. An analysis of high-speed images acquired using the LIF technique enables identification of the spatial growth rate of waves on the jet boundary, as well as the frequency of oscillation of the weakly diffusive interface. Temporal fluctuations of fluorescence intensity are found to be spatially invariant in the jet near-field, further attesting to behavior consistent with that of a self-sustained oscillation whose frequency depends on the viscosity ratio. The observed frequencies show trends similar to those of absolutely unstable modes calculated from spatio-temporal linear stability theory presented in a companion paper. Spectral Proper Orthogonal Decomposition was used to analyze the images and identify the various spatial modes, and suggests the existence of a single dominant mode. Together, these observations provide strong circumstantial evidence for the existence of a global mode that arises from the absolute instability of velocity and viscosity profiles in a region close to the nozzle exit plane.
\end{abstract}

\maketitle
\section{\label{intro} INTRODUCTION}

Shear layers with spatially variable fluid physical properties occur in a variety of industrial and natural systems. The variations in density and/or viscosity may occur due to temperature gradients, as in the case of a plasma torch \cite{Duan2002}, static mixers \cite{Cao2003}, reacting flows \cite{Pathikonda2021}, or due to species concentration gradients, such as salinity gradients set up when an estuary enters an ocean. While most situations will feature both density and viscosity effects,  the fluid dynamics of variable density jets have been extensively studied, both theoretically \cite{Huerre1985, Yu1990, Lesshafft2006} and experimentally \cite{Kyle1993, Yu1993, Hallberg2006}. In particular, low-density jets have been shown to be a member of a class of globally unstable flows \cite{Huerre1990}, which are characterized by the sudden onset of a regime with enhanced mixing, self-sustained oscillations and insensitivity to external forcing. The frequency of the global modes in the near-field of low density jets has been linked to the existence of local profiles over a finite streamwise extent that are absolutely unstable in the framework of local spatio-temporal linear stability analysis \cite{Chomaz2005, Pier2001}.  While the primary mechanism of breakdown of the flow is inviscid, arising from the baroclinic torque established by gradients in density and pressure, there are strong indications \cite{Hallberg2006} that viscosity does modify the onset of global modes, as well as their frequency. Hallberg and Strykowski \cite{Hallberg2006} conducted experiments with multiple nozzle geometries, thereby independently studying the effects of shear layer thickness, density ratio and jet Reynolds number, and   found a weak but perceptible effect of jet Reynolds number on the global mode frequency. The linear stability calculations of Lesshafft\cite{Lesshafft2006}  and Srinivasan et. al. \cite{Srinivasan2010} also suggest that the frequency and transition boundary between convective and absolute instability are affected by the viscosity in the form of the Reynolds number. It is therefore natural to inquire into the effects of variations in viscosity between the jet and the ambient medium, which is the focus of this study. \\

Strong gradients in viscosity are unlikely to be established in gas flows, and we look to other situations where the role of viscosity gradients has been more extensively investigated. In fact, in contrast to free shear flow, an extensive body of literature on variable viscosity flows addresses pressure-driven internal flows of high Schmidt number fluids \cite{Govindarajan2014b}.  While viscosity is instinctively assumed to have a stabilizing influence on the growth of disturbances, it is responsible for altering the base state of a flow, often creating sharp velocity gradients  through the no-slip condition and therefore serving as a source of disturbance kinetic energy. It has long been known that a jump in viscosity across a sharp interface can lead to long-wave instabilities at any Reynolds number ($Re$) \cite{Yih1967} or short-wave instabilities at low $Re$ \cite{Hooper1983}.  Here, we focus on flows of miscible fluids; the immiscible situation is covered in reviews by Joseph et al. \cite{Joseph1997} and more recently, Govindarajan and Sahu \cite{Govindarajan2014b}. Mention should also be made of the extensive work done on planar shear layers with gas-liquid streams \cite{ Matas2011, Otto2013, Fuster2013, Ling2019, Bozonnet2022} in the context of liquid atomization. Together, these studies have shown that viscous stability calculations are required to match theory with experiment; that absolute instability of co-flowing gas/liquid streams is supported when velocity defects immediately downstream of a splitter plate are considered, and match experimentally observed frequencies; and that confinement and the finite thickness of the gas stream play an important role in destabilization. However, viscosity ratios of the two streams, when considered in the above studies, were always extremely small and the effects of this ratio were rarely isolated.

Ern et al. \cite{Ern2003} showed the destabilizing effects of a finite thickness interface marked by gradients in velocity and viscosity, and demonstrated that for certain parameter ranges, the instability could be stronger than that of the corresponding sharp-interface configuration. Sharp gradients in velocity profile, combined with variations in viscosity profile, lead to additional source terms in the equation for disturbance kinetic energy, which drive the growth of instabilities near the diffuse interface. Ranganathan and Govindarajan \cite{Ranganathan2001} performed a temporal stability analysis of the effects of diffusion in channel flow of two fluids in a three-layer configuration, and found that when the critical layer (the region where the wave speed matches the mean velocity) overlaps the diffuse interface, the flow was strongly stabilized or destabilized depending on whether the more viscous fluid was adjacent to the wall or in the interior. Talon and Meiburg \cite{Talon2011} consider planar two- and three-layer Poisueille flows in the Stokes regime, and identify two interfacial and two bulk modes, with instability arising without a critical layer overlap. For the pipe geometry, Selvam et al. \cite{Selvam2007,Selvam2009} performed a linear stability analysis that predicted the onset of absolute instability for low Reynolds numbers when the viscosity contrast is sufficiently high, and the diffuse interface is located in a certain range of radial locations with respect to the pipe radius. They find that when the core fluid is more viscous, the flow can be at best unstable over a certain Reynolds number range, with the axisymmetric mode being dominant. When the less viscous liquid is in the core, helical modes are favored, and can lead to absolute instability.  These calculations were supported by Direct Numerical Simulations and a global linear stability analysis. Their findings partially reproduced the experimental observations of \cite{DOlce2008}, who reported axisymmetric pearl- and mushroom-shaped instabilities in the Reynolds number range of 2-60, with no observations of helical modes or sharp transitions in either wavelength or frequency that would provide strong evidence of a global mode. The helical modes observed by Cao et al. \cite{Cao2003} for injection of a low-viscosity fluid into a static mixer are in line with the theoretical results discussed above. \\

It is difficult  to translate insights from confined flows to unconfined shear layers such as jet flows due to the fundamental differences in velocity profiles, characterized by inflection points in the case of jets.  Any insights from pressure-driven flow studies have to be interpreted with caution, since confinement is known to play both stabilizing and destabilizing roles in other situations involving absolute instability of single phase \cite{Juniper2006a, Healey2009, Yang2021} or two-phase flows \cite{Bozonnet2022}. Further, the seminal works of Rayleigh \cite{Rayleigh1892} and Tomotika \cite{Tomotika1935} considered capillary flows of liquid filaments in another viscous medium  in the limit of negligible Reynolds number and are not relevant to the present work which is focused on large Reynolds numbers. Sahu and Govindarajan \cite{Sahu2014} considered a planar shear layer configuration, and the emergence of an overlap mode when the gradients in velocity and viscosity occur in the same region. Destabilization was enhanced when these layers overlapped, and with decreasing thickness of either of the gradient regions. In line with inviscid theory, the configuration was found to be absolutely unstable when countercurrent velocity profiles were used for the base state. More recently, Yang and Srinivasan \cite{Yang2024} carried out a linear stability analysis of base profiles corresponding to the near-field of a jet emerging into an ambient medium with a different viscosity. Their base profiles reflected modifications to the standard tanh- profile typically used in the analysis of jet flows \cite{Mattingly1974}, such as an inward radial shift due to the decelerating effects of a more viscous ambient medium, and concentration gradient regions that were much thinner than the momentum thickness. Similar to observations for constant-property jet flows \cite{Mattingly1974, Morris1976}, the axisymmetric and helical modes had nearly equal temporal growth rates over a wide range of conditions specified by the jet Reynolds number, ambient-to-jet viscosity ratio, momentum and concentration layer thicknesses. Further, beyond a critical value of viscosity ratio that was Reynolds number-dependent, absolute instability of the flow was supported, with the axisymmetric mode being strongly favored over the helical mode. Maharana and Sahu \cite{Maharana2023} examine plane Poiseuille flow with two fluid layers of identical viscosity that react at the interface to produce a third species of different viscosity, and contributes to the emergence of a strong Kelvin-Helmholtz instability.   

In comparison to the above numerical studies, experimental investigations of variable-viscosity free shear flows are scarce in the literature, though related studied abound. In the limit of zero inertia, miscible displacement of a more viscous liquid by a less viscous liquid and the resulting fingering instability has received significant attention. The secondary instabilities of the viscous fingers have been modeled as core-annular flows \cite{Scoffoni2001, Balasubramaniam2005}. Buoyant plumes, which exhibit a puffing instability, have been shown to be globally unstable to sinuous modes, with the Strouhal number expressing the non-dimensional  global mode frequency being a function of the Richardson number ($Ri$), which express the ratio of buoyancy to jet inertia \cite{Chakravarthy2018, Bharadwaj2017}. A recent article \cite{Pathikonda2021} examines the case of a turbulent jet of a low viscosity fluid, with annular co-flow of high  viscosity fluid in a pipe. This flow model a typical chemical mixing configuration, with high Schmidt numbers and a thin reaction zone. The viscosity gradient is shown to cause significant differences in the distribution of turbulence production, skewing towards the low viscosity fluid. 

 With the above numerical results in mind, we carry out an experimental study that seeks to isolate the effects of large viscosity contrast between a jet and its surroundings. The goals of the present study are to characterize the near-field of a low-viscosity jet at moderate Reynolds numbers ($1000 < Re < 3500$) for ambient-to-jet viscosity ratios ranging from 1 to 45, and to examine the flow field for any evidence of global modes. This  article is structured as follows. Section II describes the experimental facility used to achieve a neutrally buoyant jet with high viscosity contrast. Section III describes the flow visualization and the observation of disturbance modes. Section IV presents the measurements of velocity fluctuations and spectra obtained using hot film anemometry. Section V describes identification of the dominant modes using a Proper Orthogonal Decomposition (POD)-based technique applied to the images from visualization. Section VI provides a summary and conclusions. 

\section{Experiments}

\begin{figure}[b]
    \centering
    \includegraphics[height=4in]{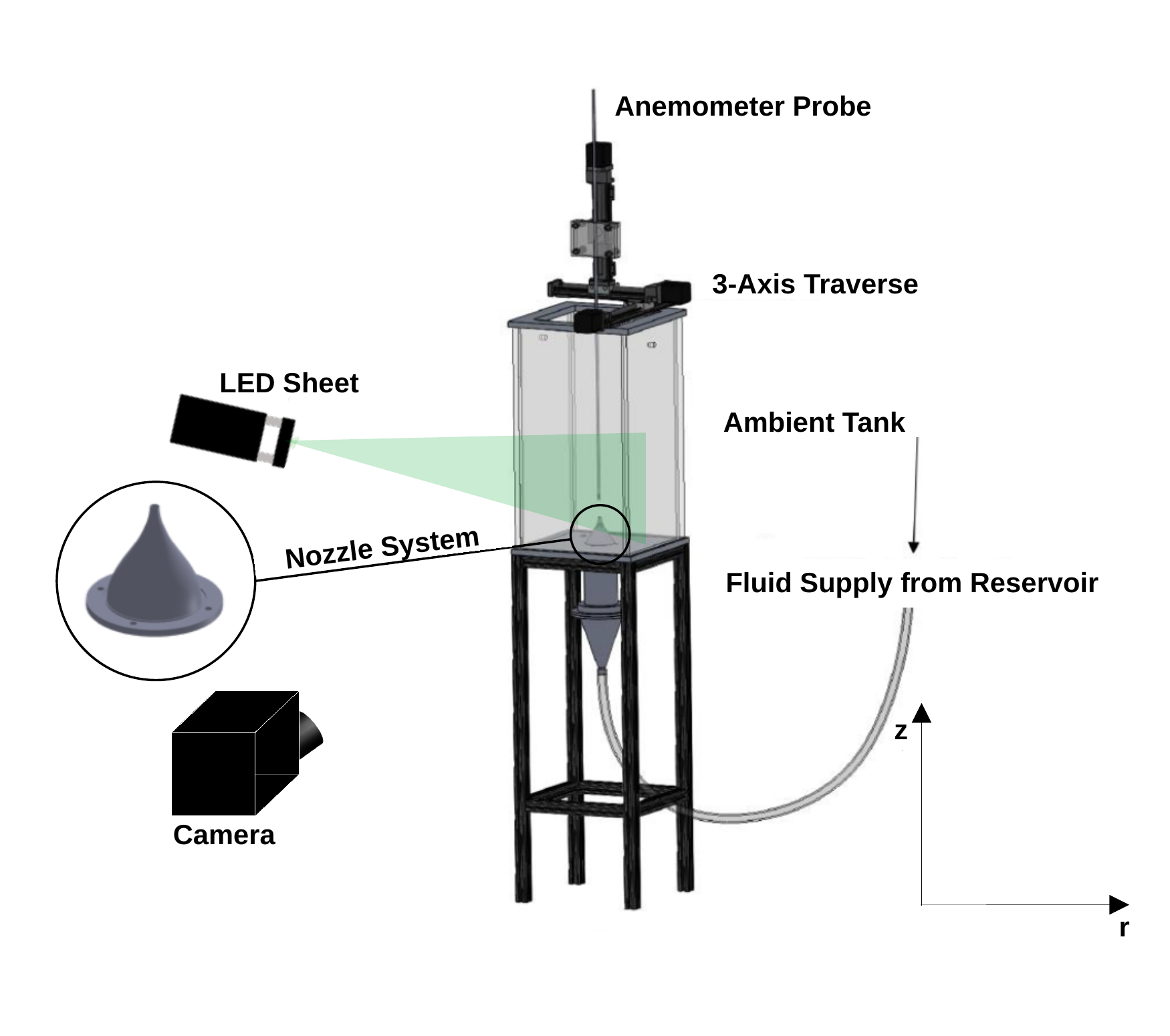}
    \caption{Sketch of the facility used to produce a low-viscosity jet using gravity-driven flow.}
    \label{fig:jetfacility}
\end{figure}

The study of the effects of viscosity gradients alone on the development of instabilities in the near-field of a jet requires the elimination of density effects, as well as a quiet facility with a minimum amount of external disturbances. Accordingly, experiments were carried out in a jet facility shown in Fig. \ref{fig:jetfacility} that utilizes gravity to attain the required flow rates. The experiments were performed in the vertical configuration. A large overhead reservoir delivered fluid to a nozzle located in a test section of square cross-section through a flow meter, a diffuser section and a flow straightener composed of laminar flow elements. The nozzle has a fifth-order polynomial profile with zero slope and curvature at its inlet and exit planes, and imposes an area contraction of 87 on the entering flow. The nozzle exit diameter $D$ is 6 mm. \\

Jet Reynolds numbers, $Re$, are defined based on the nozzle exit diameter $D$ and the average velocity $\overline{U}$ of the flow, as inferred from measurements of the volumetric flow rate from the flowmeter (accuracy of 2\%), 
\begin{equation}
    Re = \frac{\rho \overline{U}D}{\mu _j}
\end{equation}
where $\mu _j$ is the viscosity of the injected fluid. 

The requirement of having a wide range of viscosity contrast defined in terms of the ambient-to-jet viscosity ratio requires the use of liquids. Salt water of nominal density  1042 kg/m$^3$) is chosen as the test fluid, in order to facilitate density-matching as explained below. Reynolds numbers up to 4000 can be attained using this reservoir/nozzle combination. \\

The jet exhausts into the test section, which is made of transparent polycarbonate and has a square cross-section  with inner dimensions 240$\times$240 mm$^2$. Overflow ports near the top of the tank enable maintenance of a constant fluid height in the test section during operation. The top of the tank is open to allow direct mounting of a hot-film anemometry system. The fluid in the tank creates the desired  viscosity ratio, which is defined as 
\begin{equation}
    M = \frac{\mu _\infty}{\mu _j}
\end{equation}
where the subscript $\infty$ refers to test section conditions. For this study, propylene glycol and salt water were used as the two fluids. Propylene glycol in its pure form has a viscosity of 42 mPa$\cdot$s, approximately 45 times that of water, and has a density of 1036 kg/m$^3$, which is only a few percent above the density of water. Industrial-grade propylene glycol used in this work was often found to have even higher viscosity values, and therefore each batch of glycol was measured for its density and dynamic viscosity. A salt water solution was then prepared in order to match the density to within a tenth of a percent ($\frac{\Delta \rho}{\rho} = |\frac{\rho _j -\rho _\infty}{\rho _j}|  < 0.005$). These fluids are Newtonian over the range of strain rates imposed, and are very miscible with each other, eliminating surface tension as a relevant parameter. Nevertheless, as we shall see, the interface thickness has no time to develop diffusively and essentially remains a nearly sharp interface in the near-field of the jet.  

\subsection{Constant Property Jet Profiles}
\begin{figure}
\begin{subfigure}{0.48\textwidth}
    \centering
\includegraphics[height=2.5in]{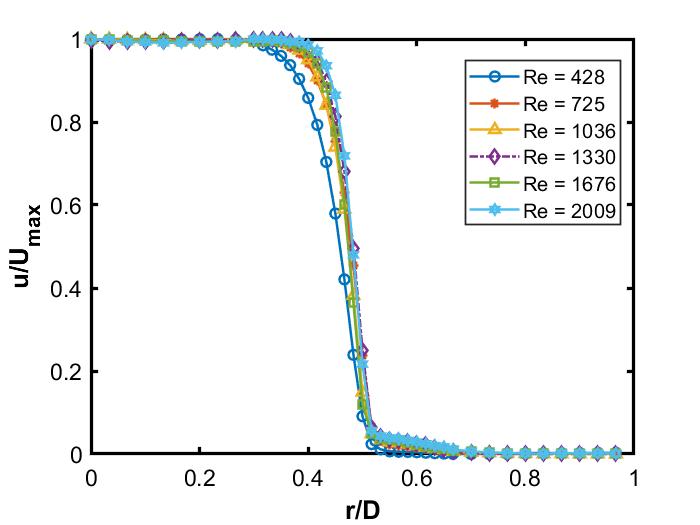}
\caption{}
\end{subfigure}
\begin{subfigure}{0.48\textwidth}
    \centering
    \includegraphics[height=2.5in]{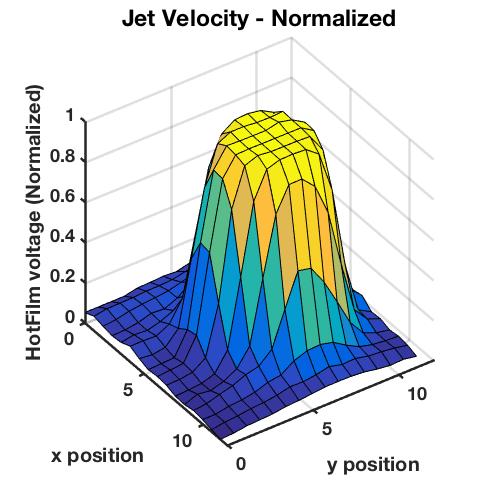}
\caption{}
\end{subfigure}\\
\begin{subfigure}{0.48\textwidth}
    \centering
    \includegraphics[width=0.95\linewidth]{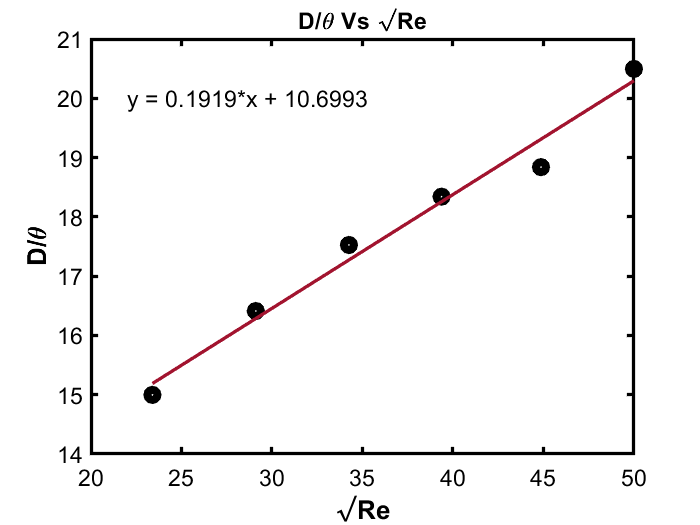}
\caption{}
\end{subfigure}
    \caption{(a) Velocity profiles at multiple Reynolds numbers. (b) Two-dimensional trace of velocity at $z/D=0.1$, showing symmetry of profile about the axis. (c) Shear layer momentum thickness as a function of Reynolds number.}    \label{fig:baseflow1}
\label{fig:meanprofiles}
\end{figure}

Hot-film anemometry was used to first characterize the jet facility to establish the base flow for a constant property jet. For a water jet issuing into a water ambient, anemometry was used to characterize the mean velocity profiles and background noise level, as well as the shear layer thickness of the jet at the nozzle exit plane. Fig. \ref{fig:meanprofiles}(a) shows velocity profiles emerging from the jet for multiple Reynolds numbers. The profiles are mostly top-hat, characterized by a steep decrease in magnitude in the shear layer towards the quiescent ambient fluid. A two-dimensional trace of voltage (Fig. \ref{fig:meanprofiles}b) at the exit plane ($z/D=0.1$) confirms the axisymmetric nature of these profiles.  Momentum thicknesses of the shear layer were evaluated as a function of Reynolds number by integrating radially from the centerline to a location where the velocity decreased to 10\% of the centerline; further radial measurements were avoided as  the hot film responds unreliably to the low velocities in the entrained flow. The momentum thickness is evaluated as

\begin{equation}
\theta =  \int_{0}^{\infty} \frac{U(r)-U_{\infty}}{U_c-U_{\infty}}\left[1-\frac{U(r)-U_{\infty}}{U_c-U_{\infty}}\right]dr
    \end{equation}

The laminar nature of the jet boundary layer at the exit plane is checked (Fig. \ref{fig:meanprofiles}(c)) by observing a linear relationship between $D/\theta$ and $\sqrt{Re}$. The constants in the fit are unique to the nozzle geometry, reflecting the acceleration imposed by the area contraction and the resultant thinning of the boundary layer entering the nozzle.  Profiles at multiple downstream locations within the first half-diameter can be well-represented by an equation of the form used by Mattingly and Chang \cite{Mattingly1974}: 

\begin{equation}
\frac{u}{U_c} = 0.5 \left[ 1+\tanh{\left(\frac{D}{8\theta}\left(\frac{1}{r}-r\right)\right)} \right]
\end{equation}

\begin{figure}
\begin{subfigure}{0.48\textwidth}
    \centering
    \includegraphics[width=0.95\textwidth]{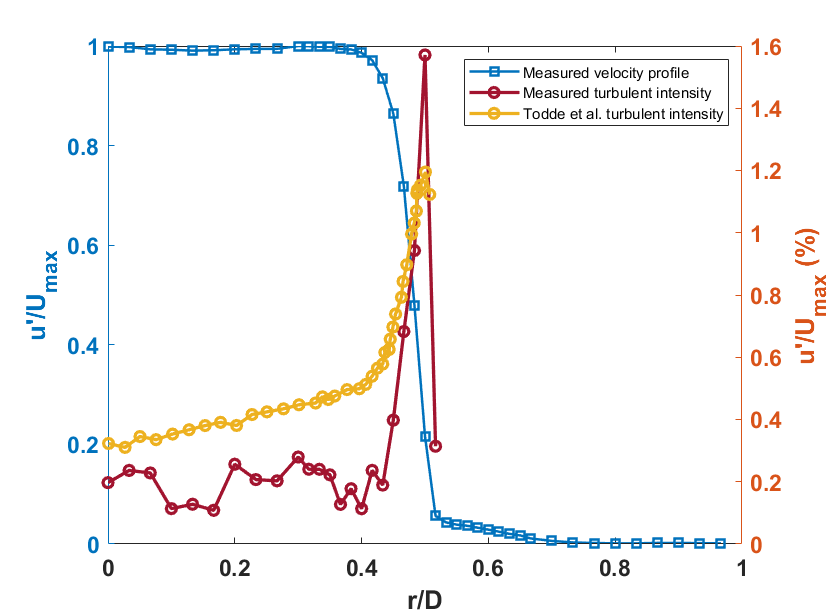}
\caption{}
\end{subfigure}
\begin{subfigure}{0.48\textwidth}
    \centering
    \includegraphics[width=0.95\linewidth]{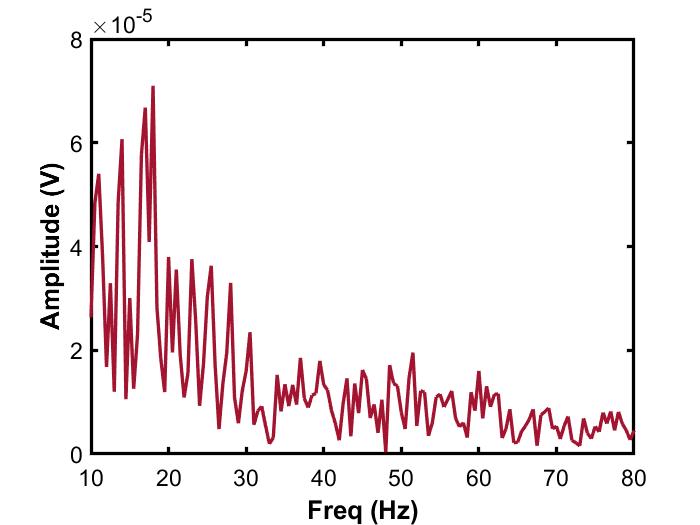}
    \caption{}
\end{subfigure}\\
    \caption{(a) Radial profile of turbulence intensity at $Re = 1688, M=1$. (b) Frequency spectrum of voltage fluctuations.}    \label{fig:turb_base}
\end{figure}

The turbulence characteristics of the flow at the exit plane are now presented. The profiles of normalized turbulence intensity are shown in Fig. \ref{fig:turb_base}(a) alongside profiles measured by Todde et al.(2009)\cite{Todde2009} in their work on low Reynolds number free jets. It can be seen that the turbulence intensity profile prevailing in the present facility has a comparable trend, although it has lower centerline turbulence intensity. This speaks to the benefit of having a gravity-fed jet, free from any influences of upstream pumps.  Lastly, we  examine the spectral content of the flow at the exit plane in Fig. \ref{fig:turb_base}(b), and find no discrete peaks in the frequency spectrum, assuring that the jet is a low-turbulence system with little ambient noise.

\subsection{Test Procedure}
For experiments with propylene glycol as the ambient fluid, each set of experiments is conducted at a constant $Re$ governed by the flow rate. Over the course of several test runs, the viscosity of the tank (and hence the value of $M$) decreases due to mixing with the injected salt water. This makes the test runs inherently quasi-transient. Therefore, a careful procedure was followed to minimize the effects of variation of $M$ during each test run. We first estimate the variation of $M$ during a typical test run. At a salt water-glycol interface, diffusion acts to thicken the interface to yield a concentration thickness given by $\sqrt{\gamma t} $   where $\gamma$ is the binary diffusion coefficient of propylene glycol into water ($1.1 \times 10^{-9}$ m$^2$/s). For a one-minute long trial run, this yields a diffusion length of the order of $0.01D$. In practice, test runs were much shorter, typically lasting 20-30 seconds after the initial starting vortex had passed out of the field of view. During this period, high-speed images were acquired with a digital camera operating at 500 fps and at $1024\times1024$ pixels resolution. After the image acquisition was completed, the flow was turned off, the tank was stirred with a mixer and allowed to settle and become quiescent again, before the next trial (typically 30 minutes). A sample of tank fluid was taken for subsequent viscosity and density measurements for determining the value of $M$ for each trial. In this way, at each $Re$, values of $M$ starting from 50 and descending down to 15 were attained. \\

\section{Results}
\subsection{Flow Visualization}
\begin{figure}
    \centering
    \begin{subfigure}{\textwidth}
    \centering
    \includegraphics[width=0.2\textwidth]{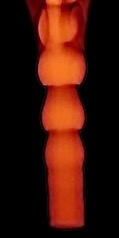}
    \includegraphics[width=0.2\textwidth]{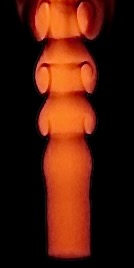}
    \includegraphics[width=0.2\textwidth]{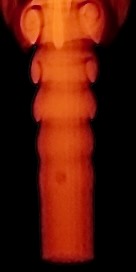}
    \includegraphics[width=0.2\textwidth]{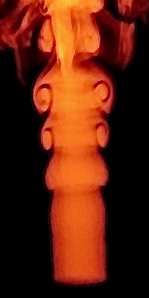}
    \caption{}
    \end{subfigure}
    \centering
    \begin{subfigure}{\textwidth}
    \includegraphics[height=3.5in]{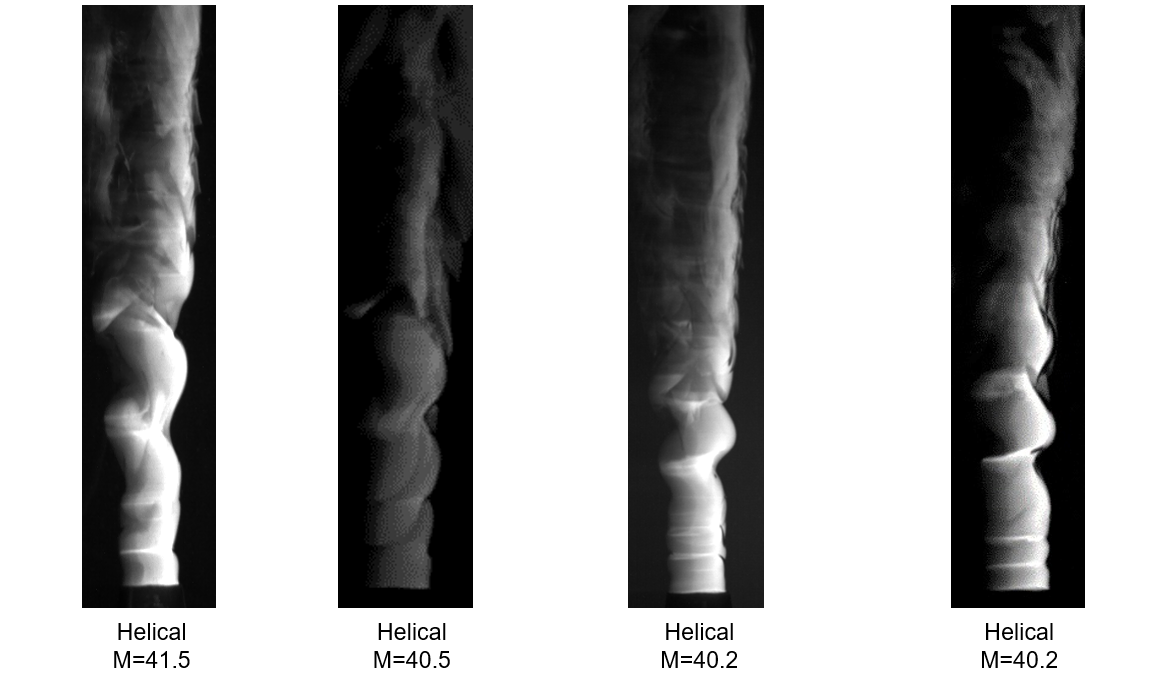}
    \caption{}
    \end{subfigure}
    \caption{(a) Growth of axisymmetric instabilities for $M=1$ and multiple $Re$ (from left to right) $Re= 1036$, 1546, 2009 and 2540, (b) Helical modes observed at $M\sim41$ for (left  to right) nominal values of $Re=1600$, 1800, 2000 and 2400.}
    \label{fig:flowviz}
\end{figure}

Preliminary images for $M=1$ (water jet into water ambient) were acquired with an 18MP camera whose lens was equipped with an orange filter. The jet fluid was dyed with Rhodamine 6G, and  the tank volume was illuminated with a blue LED light. The emission by rhodamine in the orange part of the spectrum was captured and shows the breakdown of the jet. Fig. \ref{fig:flowviz}(a) shows the axisymmetric nature of the instabilities dominating the breakdown process, after developing from an initial nearly parallel near-field region. As $Re$ is increased from approximately 1000 to 2500, this distance over which the jet retains a coherent character becomes palpably shorter. Unlike the observations of Mattingly \& Chang \cite{Mattingly1974}, no evidence of an eventual competition between the axisymmetric mode and a growing helical mode in the far-field is observed, and the images show the classic vortex roll-up process associated with the Kelvin-Helmholtz instability. On the other hand, when the jet emerges into an ambient medium of propylene glycol ($M\approx 41)$, helical instabilities are observed over a range of $Re$ from approximately 1600 to 2400, as seen in Fig. \ref{fig:flowviz}(b). Of note is the disappearance of the parallel flow region in the near-field, with the helical mode almost instantaneously developing at the exit. We also note that the wavelength of the disturbances seems substantially lower than that of the axisymmetric instability at $M=1$. \\

\begin{figure}
    \centering
    \includegraphics[height=4in]{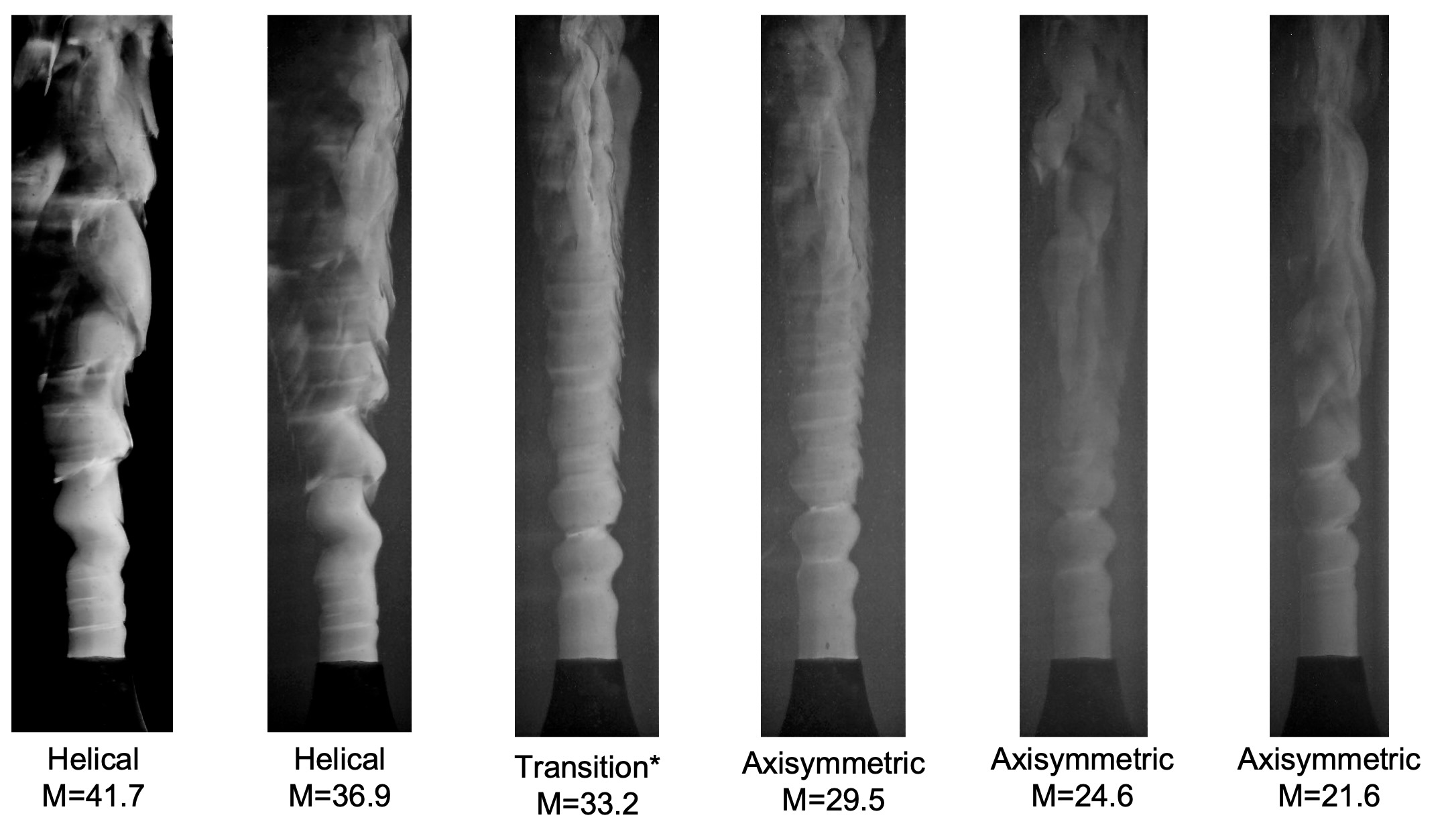}
    \caption{Images showing the transition from helical to axisymmetric modes as $M$ is decreased, for $Re=2400$.}
    \label{fig:const_Re_varying_M}
\end{figure}

\begin{figure}
    \centering
    \includegraphics[height=3in]{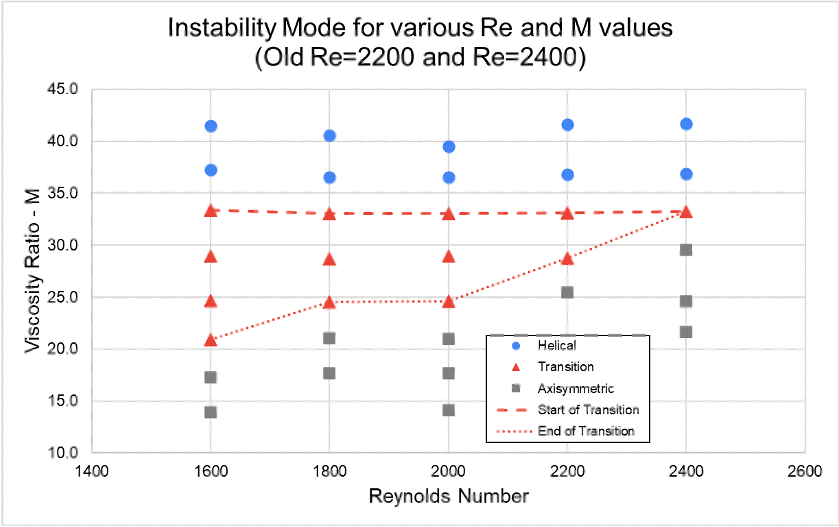}
    \caption{Transition boundary between the helical and axisymmetric modes in ($M$, $Re$) space.}
    \label{fig:M_Re_plane}
\end{figure}

These two sets of images suggest that there must exist a transition value (or range) of $M$ for every fixed value of $Re$, where the dominant mode changes from axisymmetric to helical, and experiments were conducted to elucidate the transition behavior. Fig. \ref{fig:const_Re_varying_M} shows a sequence of images captured for $Re=1600$. The transition of the dominant instability from helical to axisymmetric, as $M$ decreases from 41.17 to 21.6 is clearly evident. Nevertheless, it is difficult to assign a precise value for the transition value of $M$ with confidence in all cases.  
Due to the nature of the experiment, involving discrete steps in $M$, a fine-grained transition value could not be determined in all cases. However, observations clearly indicate that this transition value of $M$ is $Re$-dependent. Therefore, we define a transition zone of M whose extreme values are characterized by  distinctive axisymmetric or helical modes. Fig .\ref{fig:M_Re_plane} shows our estimate for the transition zone of $M$ as a function of $Re$. Below $Re=1600$, the instability was weak, and it was difficult to distinguish the nature of the mode. For large values of $M$ that lie outside the transition zone, the instability was observed to be clearly helical (filled circles) while for low values of $M$, it was clearly axisymmetric (filled squares). The transition zone observations are marked by filled triangles. As $Re$ increases, the width of the transition zone (in terms of the range of $M$) decreases. The decrease in width of the transition zone is mostly because of the increase in the lower bounding value of $M$ below which the instability is clearly axisymmetric. At $Re=2800$ (not shown in Fig. \ref{fig:M_Re_plane}, the first trial showed a helical mode before switching into an axisymmetric mode, hinting at a potential crossover in modes at higher values of $Re$ and $M$.

\clearpage
\section{Hot Film Anemometry}
\begin{figure}[hb]
    \centering
   \includegraphics[width=6.5in]{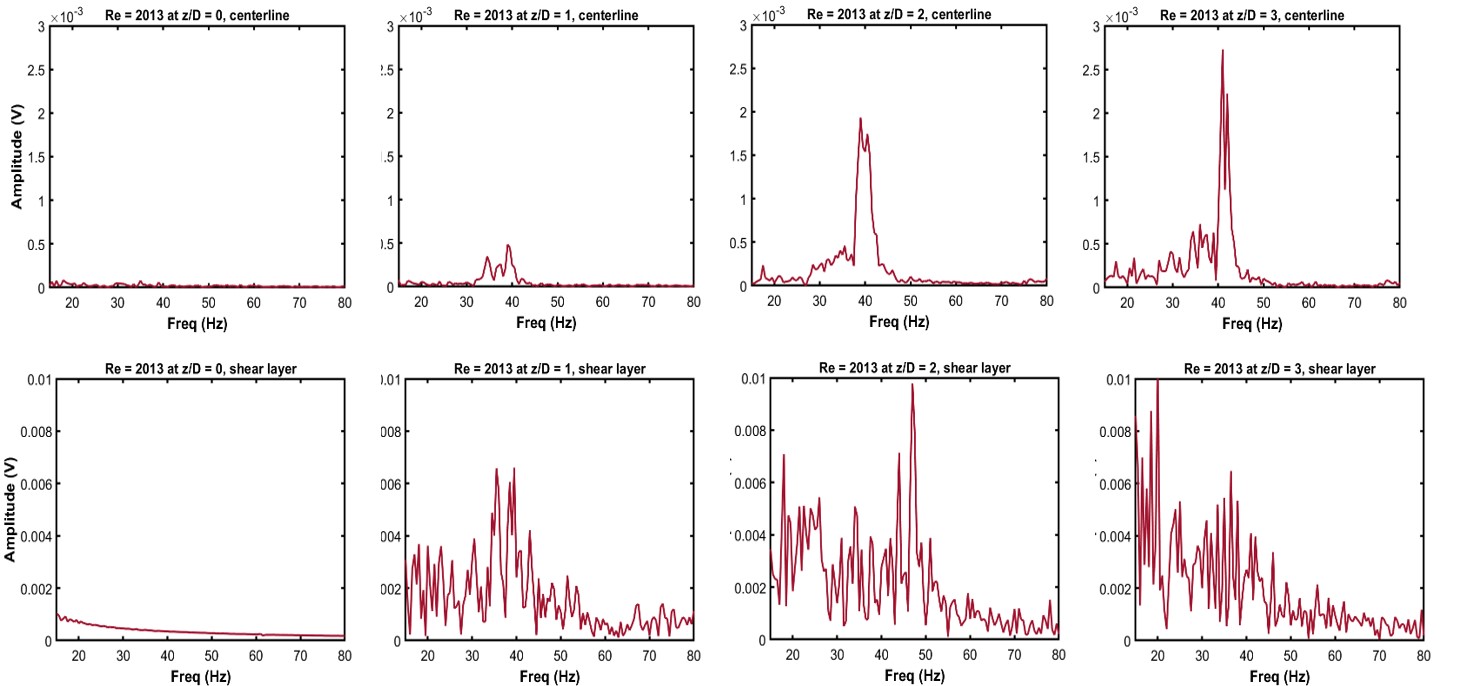}
    \caption{Evolution of velocity spectra in the downstream direction for $M=39$, $Re=2013$  on the jet axis and in the shear layer. Top row: centerline variation. Bottom row: spectra in the shear layer.} 
    \label{fig:HW_FFT_vs_z}
\end{figure}

Hot film anemometry was used to characterize the flow for values of $M$ greater than unity. The probe was originally calibrated in water, while the jet shear layer has an unknown concentration of salt water and glycol and therefore a variable Prandtl number. Since the original calibration can no longer be used, we focus on the voltage signal. Here, we are interested in the spectral content of the velocity fluctuations at different downstream distances, as well as the rate of growth of the disturbance magnitude relative to the constant property jet. Fig. \ref{fig:HW_FFT_vs_z} shows the evolution of the spectrum along the centerline and in the shear layer for $M=39$ and $Re=2013$ for $z/D=0.1$, 1, 2 and 3. The peaks in the centerline and shear layer signals are located at the same frequency, though the shear layer signal is noisier and of greater magnitude. The frequency value also appears unchanged at different downstream locations ranging from $z/D=0.01$ to $z/D=3$. A distinct frequency peak is visible at all locations in the near field, though the fixed range on the abscissa renders the signal at $z/D=0.1$ nearly invisible. 

\begin{figure}
\begin{subfigure}{0.49\textwidth}
 \includegraphics[width=\textwidth]{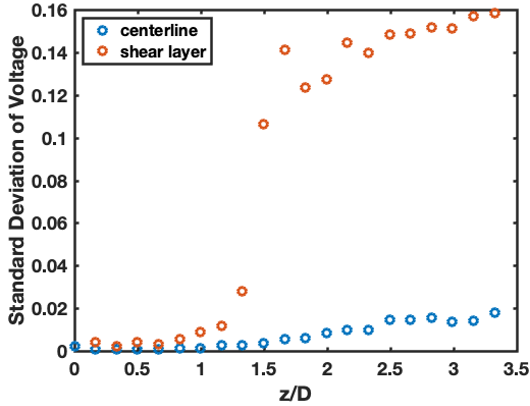}   
\caption{}
\end{subfigure}
\begin{subfigure}{0.49\textwidth}
\includegraphics[height=2.5in]{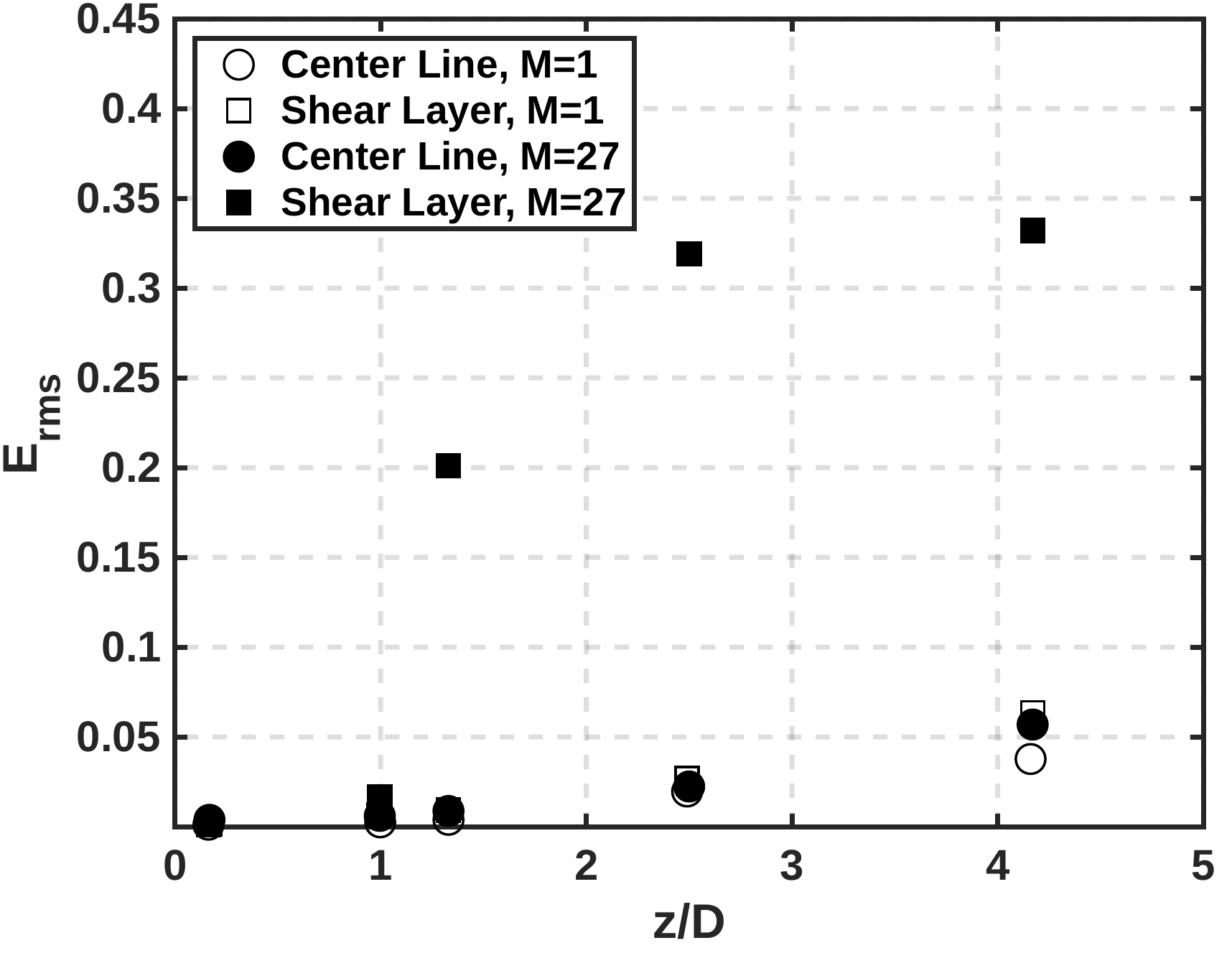}
\caption{}
\end{subfigure}
\caption{Root mean square value of voltage fluctuations along the centerline and shear layer for (a) $M=11$, $Re=1682$, and (b) $M=45$ and $M=1$ at $Re=1682$.}
\label{fig:vol_fluc_downstream}
\end{figure}

The viscosity of glycol is initially a steeply decreasing function of dilution, before decreasing gently at lower $M$ values. This meant that performing runs at large values of $M$ required large amounts of glycol. Therefore, initial runs characterizing the evolution of disturbance kinetic energy were carried out at a low value of $M=11$, which corresponds to the axisymmetric mode. Fig. \ref{fig:vol_fluc_downstream}(a) shows the evolution of the root-mean-square amplitude of voltage fluctuations along the jet centerline and in the shear layer for $M=11$, $Re= 1682$. The voltage fluctuations at the centerline show a weak increase in the downstream distance, however the fluctuation magnitude in the shear  layer follow a dramatically different trend,  starting to rise exponentially at around $z/D=1.2$ before nearly saturating at $z/D=2.5$ with a value that is nearly an order of magnitude higher than at  the exit plane.  These measurements were performed with step-wise increments in the downstream direction of $z/D$ as low as 0.16. From the observed evolution of voltage fluctuations, five points were selected for investigation for the case of large $M$ ($M\approx45$). The corresponding voltage fluctuation amplitude is shown in Fig. \ref{fig:vol_fluc_downstream}(b). A behavior similar to the case of $M=11$ is observed, characterized by an exponential increase in disturbance amplitude starting at $z/D=1$ and saturating near $D=4$. Of note is the much higher value of fluctuation amplitude at any location for $M=45$ compared to the case of $M=11$. This second case of $M=45$ corresponds to the helical mode. For reference, Fig. \ref{fig:vol_fluc_downstream}(b) also plots the voltage fluctuation behavior for $M=1$ which shows a very weak increase in disturbance energy when plotted on this scale.  

\begin{figure}
\centering
\includegraphics[height=3in]{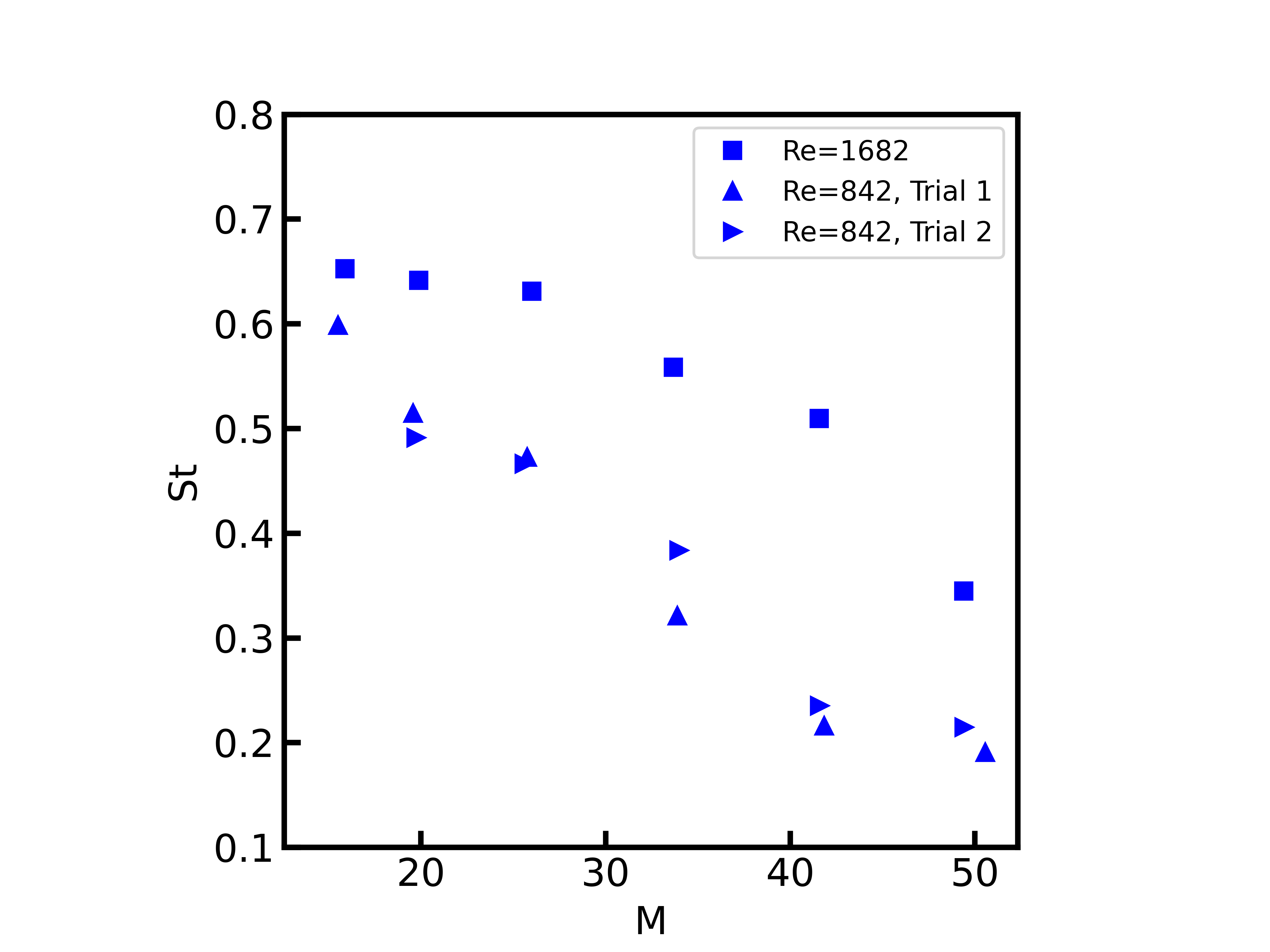}
\caption{Strouhal number, $St$, corresponding to the dominant frequency, as identified by hot film anemometry, as a function of $M$ for $Re=842$ and 1682.}
\label{fig:HW_f_vs_M}
\end{figure}

Noting that the dominant frequency did not change value along the centerline in the near-field of the jet, a series of runs at constant $Re$ and decreasing $M$ were carried out. The dependence of the dominant frequency on the viscosity ratio, as detected by hot film anemometry, is plotted in Fig. \ref{fig:HW_f_vs_M} in terms of the Strouhal number ($St=\frac{fD}{\bar{U}}$) for two values of $Re$, $Re=841$ and 1682. It can be seen from the trials for $Re=841$ that the data are fairly repeatable.  Following the experimental sequence and moving from high values of $M$ to low values, one sees an increasing trend while the helical mode remains dominant. A very interesting result from this plot is that if one follows the experimental sequence and goes from large $M$ to small $M$, one would expect to observe a change in the characteristic frequency of the disturbance as the mode switches from helical to axisymmetric. Yet, no such sharp break in behavior is apparent. The data follow a smooth trend, rising from low values of $St$ at large $M$ to high values at relatively lower $M$.

\subsection{Image Analysis}
\begin{figure}[t]
    \centering
   \includegraphics[height=2.5in]{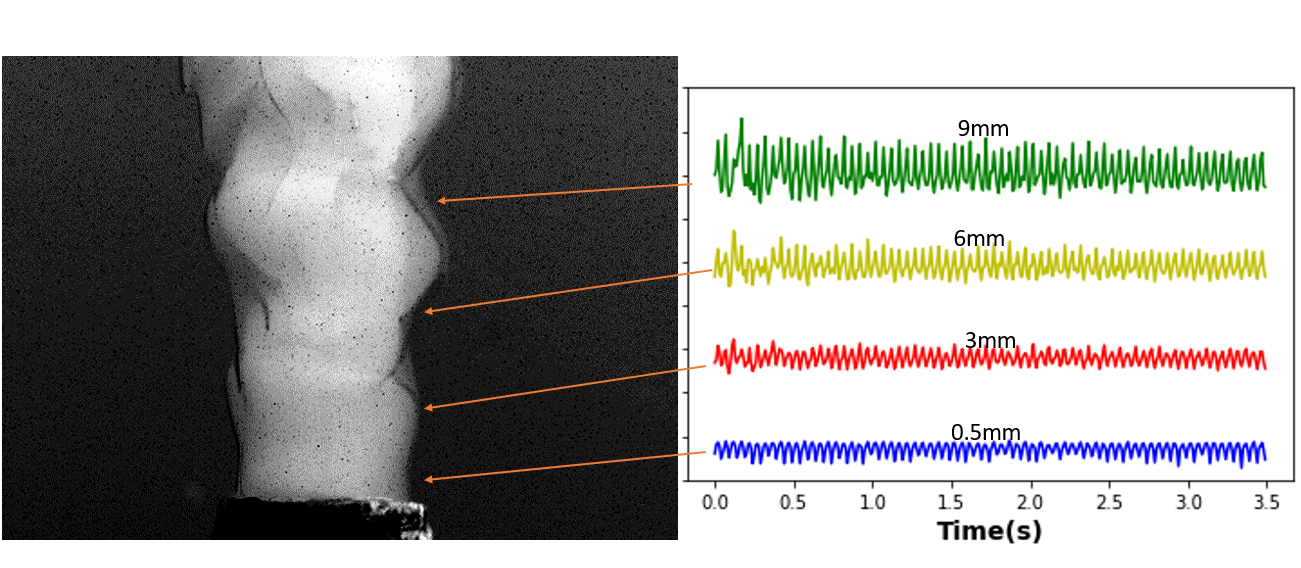}
    \caption{Temporal behavior of the jet width in pixels at different downstream distances, suggesting a single harmonic oscillation.}
    \label{fig:jet_width_oscillations}
\end{figure}

\begin{figure}
    \centering
    \begin{subfigure}{0.49\textwidth}
    \includegraphics[width=0.95\textwidth]{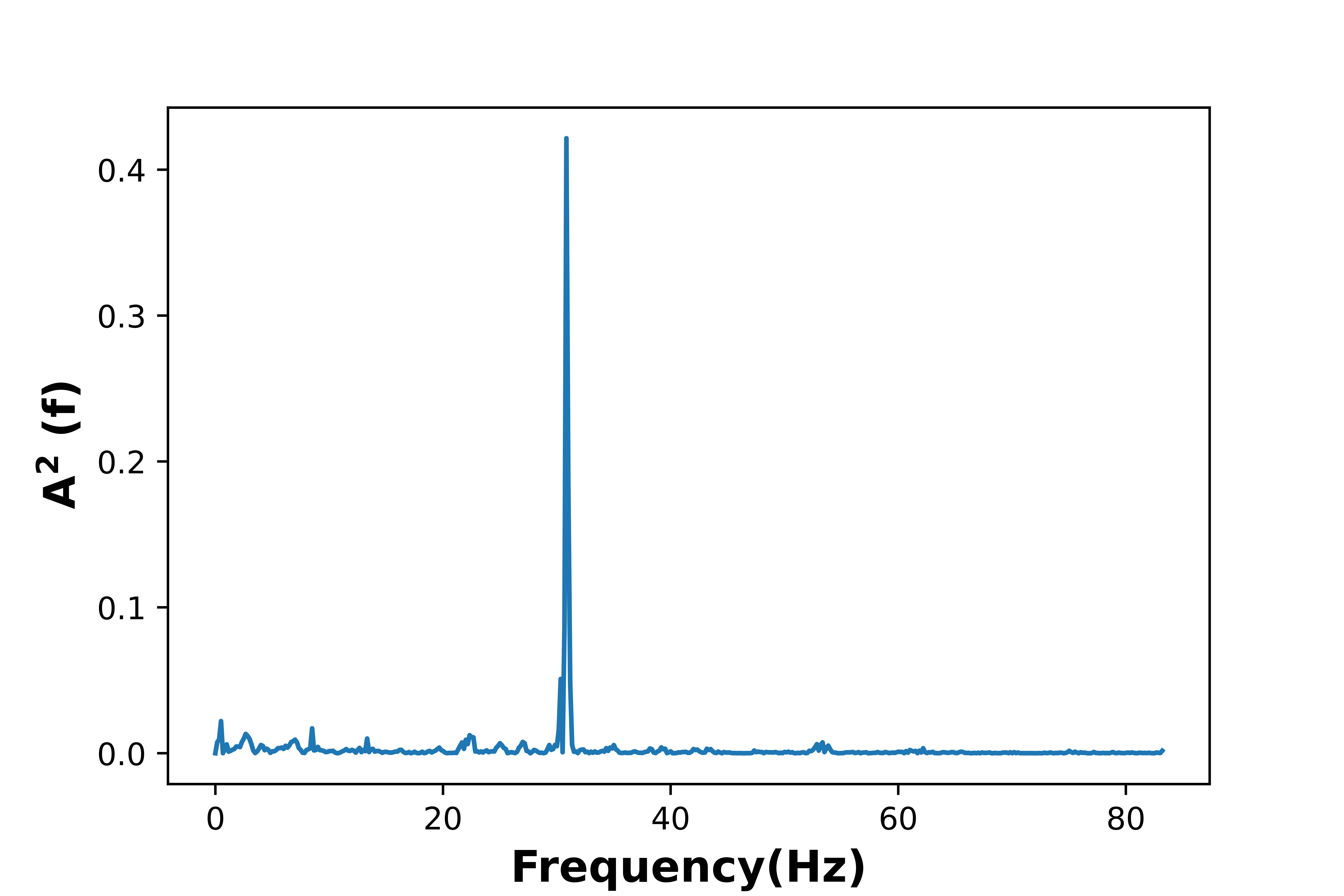}
    \caption{}
    \end{subfigure}
    \begin{subfigure}{0.49\textwidth}
    \includegraphics[width=0.95\textwidth]{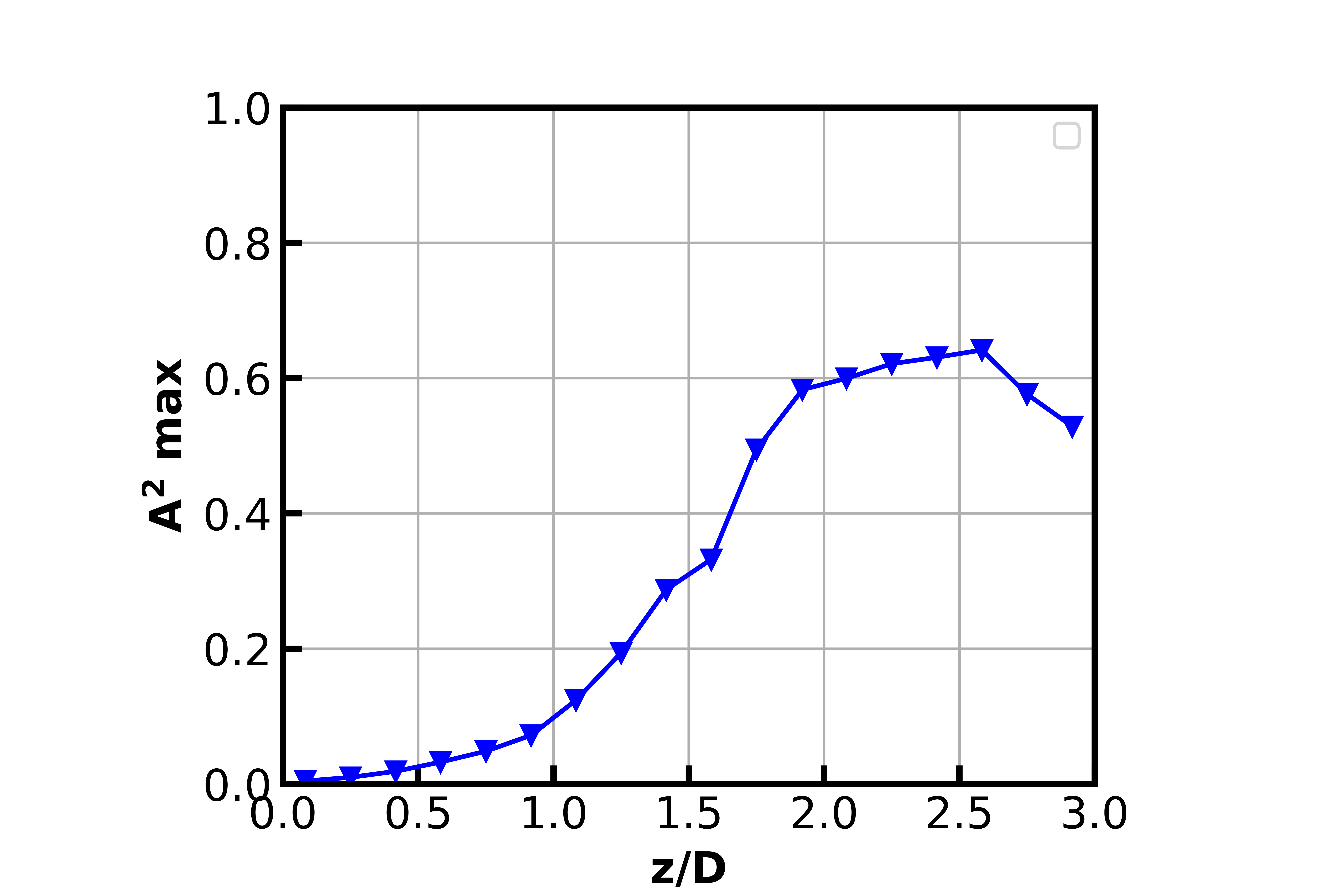}
    \caption{}
    \end{subfigure}
    \caption{(a) Power spectrum of oscillation of jet width at $z/D=1$ for $M=38$, $Re=2400$. (b) Growth of the square of the amplitude of the Fourier coefficient of the dominant mode in the downstream direction.}
    \label{fig:interface_FFT}
\end{figure}

The hot film measurements  strongly indicate the existence of a single dominant mode that saturates in intensity in the first few diameters downstream of the jet exit. However, the increased conductivity of the liquid due to the dissolved salts resulted in increased contamination, pickup of electrical line noise despite probe shielding, and the occasional air bubbles introduced into the tank due to mixing between trials that would stick to the hot film. Together, these effects resulted in a very low rate at which meaningful data were acquired. As a result, Laser-Induced Fluorescence (LIF) data were chosen as a means of investigating the growth of unstable modes. The orange filter on the camera lens ensured that the jet could be strongly distinguished against the background, by isolating the emission from the Rhodamine dye under blue illumination. Applying a threshold intensity to the grayscale images allows determination of the jet boundary. The diameter of the jet $W(z)$ at any axial location $z$ is determined from the pixel locations where the grayscale value shows a steep decline. The result was very weakly sensitive to the threshold value, but we are primarily interested in the frequency, which is unaffected by the choice of threshold. To study the spatial evolution of the oscillations of the interface, we examine the jet width at 4 locations downstream of the jet exit, as plotted in Fig. \ref{fig:jet_width_oscillations}. The jet width at each of these locations is seen to have a temporally oscillating behavior, with the amplitude of oscillations indicating a non-linear increase. In Fig. \ref{fig:interface_FFT} we further examine the amplitude and frequency of these oscillations. These oscillations, modeled as traveling waves can be decomposed into Fourier modes given by $A(z,f)e^{i(\phi z-2\pi ft)}$. Fig. \ref{fig:interface_FFT} shows the sharp peak in the power spectrum that might be surmised from the oscillatory waveform in Fig. \ref{fig:jet_width_oscillations}. Fig. \ref{fig:interface_FFT}(b) shows the variation of the amplitude of the dominant frequency in the downstream direction. Again, as with the anemometry measurements, the oscillations show an exponential increase in the disturbance amplitude with downstream distance, before saturating at $z/D=2$.

\begin{figure}[ht]
    \centering
    \includegraphics[width=\textwidth]{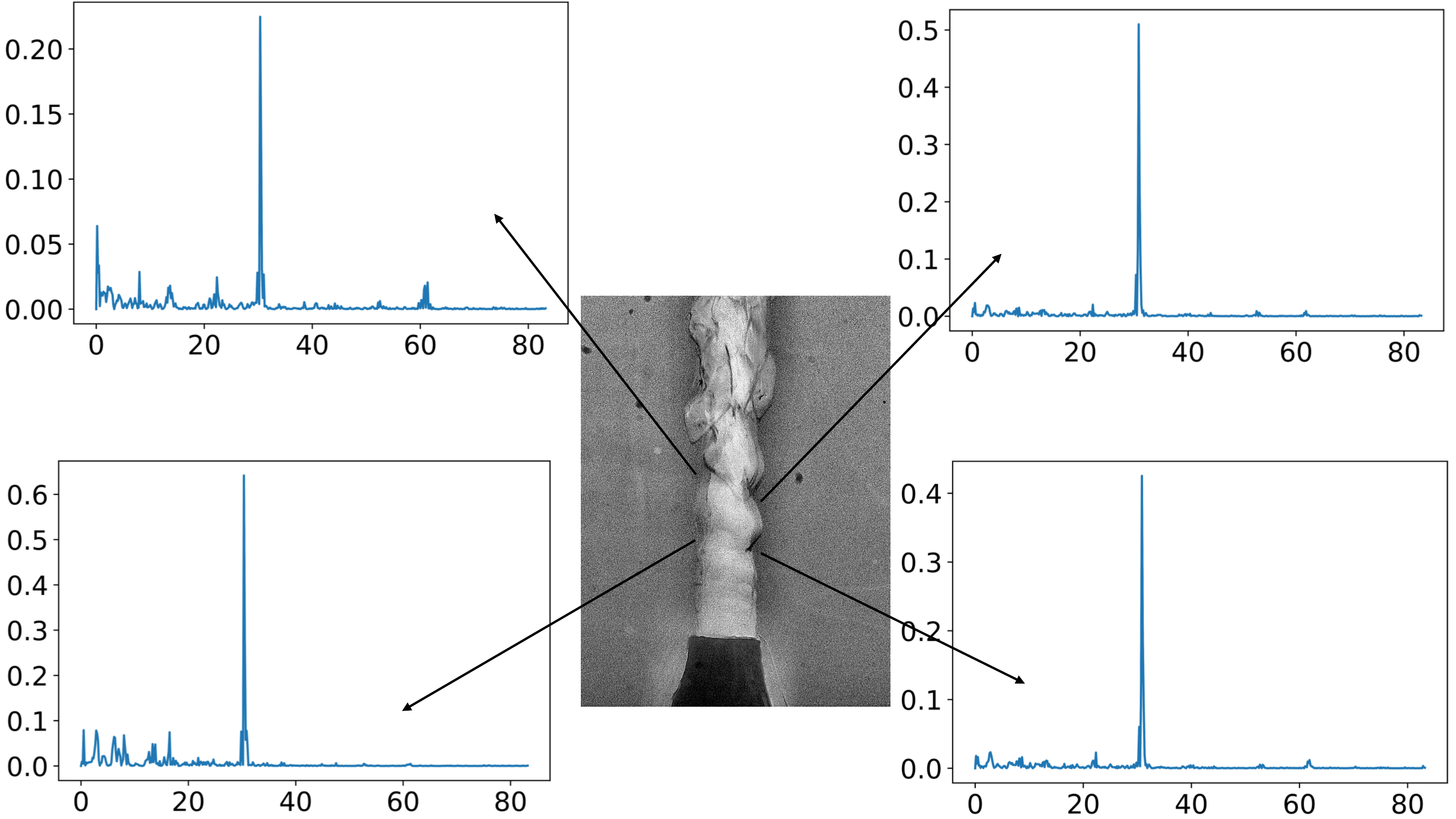}
    \caption{Spectrum of pixel intensity fluctuation in shear layer at four locations in the near-field of the jet, for $Re=2400$ and $M=38$.}
    \label{fig:pxfluc}
\end{figure}

To ascertain the nature of  this instability, that develops much faster than the axisymmetric instability of the constant property jet, we verified that the frequencies at the different downstream stations shown in Fig. \ref{fig:jet_width_oscillations} are identical. Another way of assessing the spatially invariant `global' nature of this frequency is to examine the intensity records of a single pixel in the shear layer. Fig. \ref{fig:pxfluc} shows the frequency spectrum of 4 pixels at two different downstream locations, on either side of the jet, picked carefully to ensure that they reflected the fluctuations in grayscale intensity as the wave crests move downstream. The frequencies are identical, and provide further circumstantial evidence that the instability observed at large $M$ is a global mode, likely corresponding to absolute instability of the near-field profiles.  This putative global mode has a frequency that depends on the parameters that define the flow, such as the inlet Reynolds number, the Schmidt number (the ratio of species and momentum diffusivity), viscosity ratio $M$, and the inlet velocity profile, specified by the momentum thickness $\theta$. In the experiment, the values of $Re$ and $\theta$ are conjoined through the specific geometry of the nozzle. Further, it is experimentally difficult to conduct trials at constant $M$, while changing $Re$. Therefore, we present the global mode frequency at constant $Re$ (and $\theta$) as a function of $M$.\\

\begin{figure}[ht]
    \centering
    \begin{subfigure}{0.49\textwidth}
    \includegraphics[width=0.9\textwidth]{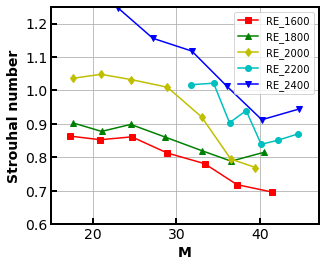}
    \caption{}
    \end{subfigure}
    \begin{subfigure}{0.49\textwidth}
    \centering
    \includegraphics[width=0.9\textwidth]{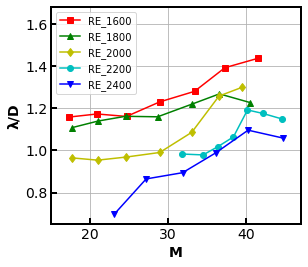}
    \caption{}    
    \end{subfigure}
 \begin{subfigure}{0.49\textwidth}
   \centering
    \includegraphics[width=0.9\textwidth]{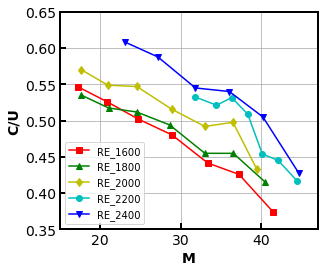}
    \caption{}
    \end{subfigure}
    \caption{Variation of instability characteristics along the constant ($Re$, $\theta$) curve, as $M$ decreases during the experiment. (a) Frequency, (b) Wavelength, and (c) Phase velocity.}
    \label{fig:f_lambda_vs_M}
\end{figure}

Therefore, a set of additional runs were carried out solely for the purpose of image analysis at $Re$ close to 1600, 1800, 2000, 2200 and 2400. Fig. \ref{fig:f_lambda_vs_M}(a) presents Strouhal numbers as a function of $M$ for these $Re$, where the dominant frequencies are calculated using temporal records of single pixels in the shear layer. The qualitative trend is similar to that observed for the hot film anemometry results for lower $Re$, going from low $St$ at large $M$ to larger values at small $M$ and saturating at the smallest values of $M$ covered. Of note are the downward jumps in $St$ as one moves leftward on the plot from the largest values of $M$ for $Re=2400$. This jump also corresponds to the visually observed change in the instability from strongly helical to a transitional behavior. Similar jumps can be observed to a lesser degree for $Re=2200$ and 2000, though not for the lower values of $Re$. The corresponding wavelengths, as determined from inspection of the images, are shown in Fig. \ref{fig:f_lambda_vs_M}(b). Knowledge of the wavelength and frequency allows us to calculate the phase velocity of the dominant mode; this is plotted in Fig. \ref{fig:f_lambda_vs_M}(c).

\begin{figure}[h]
    \centering
    \includegraphics[height=2.5in]{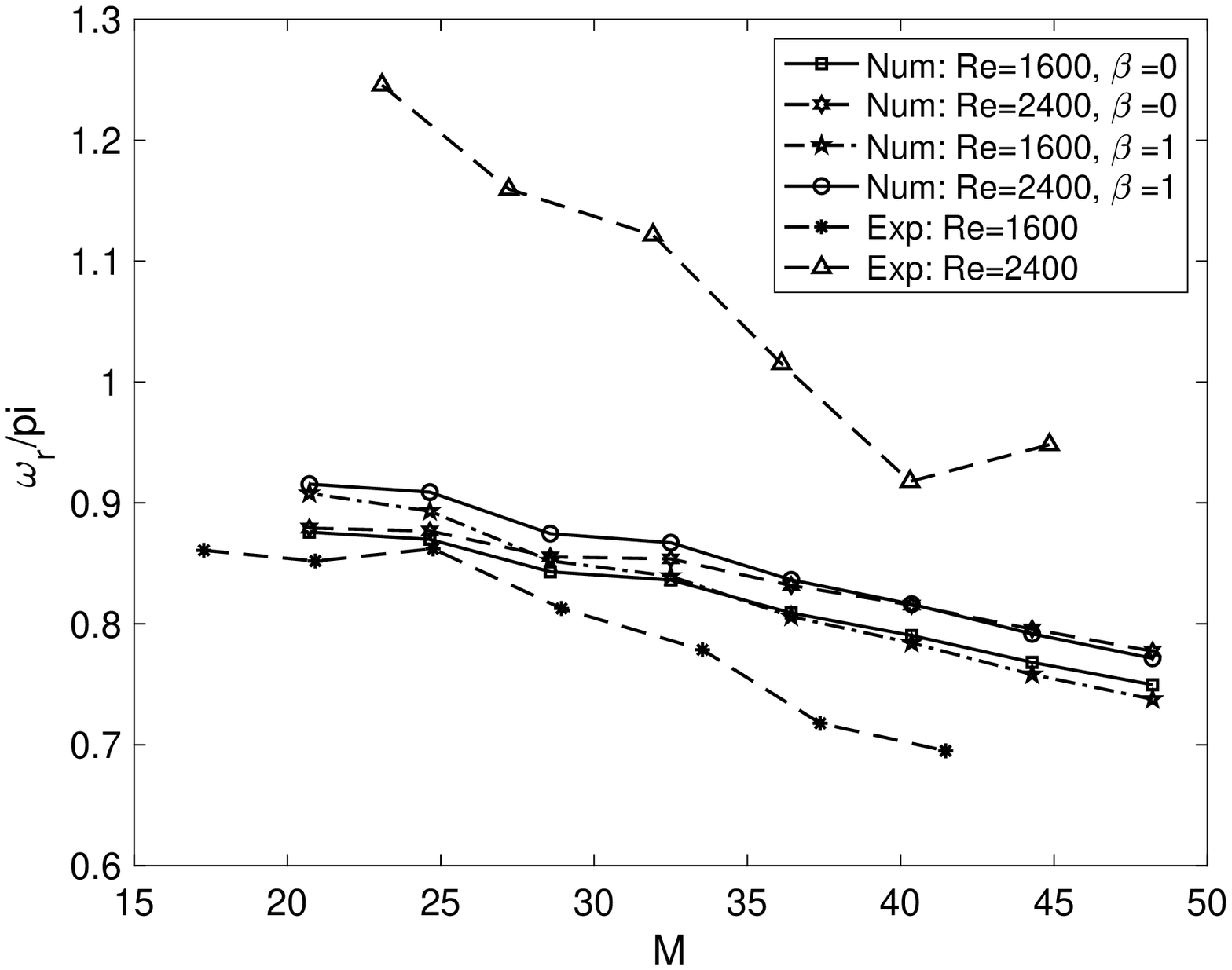}
    \caption{Comparison of the observed dependence of Strouhal number on $M$ for $Re=1600$ and $2400$ with the predictions of the absolutely unstable frequencies using spatio-temporal linear stability analysis from \cite{Yang2024}.}
    \label{fig:St_compare_theory}
\end{figure}
In Fig. \ref{fig:St_compare_theory}, the observed Strouhal numbers are compared with the predictions of linear stability analysis. The hypothesis is that the exponential growth of the experimentally observed instability arises from a linear mechanism whose origins lie in the extreme near-field of the jet close to the exit plane. Therefore, it is likely that the spatially invariant frequencies observed for a few diameters downstream correspond to the absolute instability of local profiles in the jet near-field. As mentioned in Section I, Yang and Srinivasan \cite{Yang2024} recently performed linear stability calculations for representative jet velocity profile with assumed viscosity profiles, and found a range of ($Re$, $M$, $\theta$, $\theta _\mu$) that support absolute instability of both helical and axisymmetric modes. Here, $\theta _\mu$ refers to the region over which the viscosity gradients are significant. For comparison with experimental results, the values of $Re$ and $M$ are taken from experiment and an extremely small value of $\theta _\mu$ is assumed, reflecting the negligible time allowed for the diffusive concentration interface to establish itself. A diffusive calculation for the interface thickness at $z=D$, for example, yields $\theta _\mu /D\sim 5/\sqrt{Pe}$ where $Pe = \frac{\overline{U}D}{\gamma} =ReSc$ is the Peclet number, $\gamma =1.06\times 10 ^{-9}$m$^2$/s is the species diffusivity corresponding to binary diffusion of propylene glycol into salt water \cite{wang2010binary}, and $Sc$ is the ratio of momentum to species diffusivity, with values $\sim 1000$. This yields $\theta _\mu/D= 0.011$, and is likely to be considerably smaller closer to the nozzle exit plane. Fig. \ref{fig:St_compare_theory} compares the measured and predicted trends for $St$ as a function of $M$ for two values of $Re$, $Re=1600$ and 2200 for an assumed viscosity stratification thickness of $\theta _\mu/D =0.0025$. The modeling effort assumes thin boundary layers and a similarity-based velocity profile obtained by solving a nonlinear Blasius-type boundary layer equation with variable viscosity, with the viscosity stratification thickness as an input parameter from which velocity boundary layer thickness $\theta$ is calculated to yield the base state for perturbation analysis. The trend of $St$ vs. $M$ is captured reasonably well, though the magnitude is off by a factor of 2 for $Re=2400$. This may reflect the poor understanding in the modeling approach of the details of the non-similar profiles of velocity and viscosity near the jet exit plane. The predicted frequencies show little dependence on Re, while the experimental values show a substantial dependence. This also reflects the important effect in the experimental situation, wherein the boundary layer thickness is not an independent parameter but is linked to the Reynolds number through the nozzle geometry, which is not accounted for in the similarity-based generation of mean profiles for stability analysis

\begin{figure}
    \centering
    \includegraphics[height=2.5in]{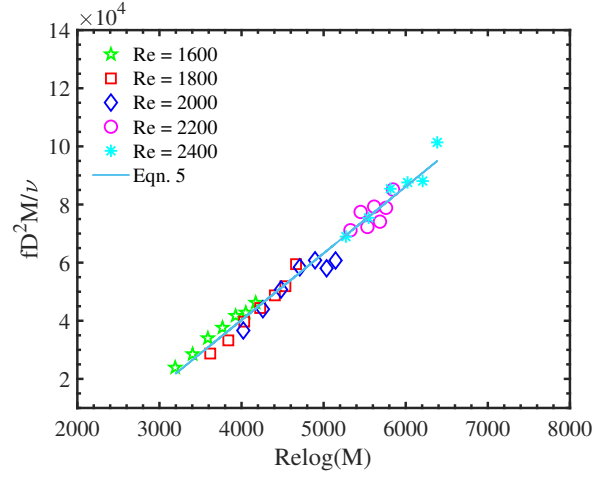}
     \caption{Non-dimensional frequency as a function of Reynolds number and viscosity ratio.}
    \label{fig:universality}
\end{figure}

We further explore whether the observed frequencies, expressed as a Strouhal number in Fig. \ref{fig:f_lambda_vs_M} for different $Re$, can be collapsed onto a single universal curve with appropriate choice of non-dimensional coordinates. The plot of Strouhal number clearly shows the inadequacy of using an inertial time scale to collapse the data, and therefore we seek to non-dimensionalize using the viscous time scale $D^2/\nu$ where $\nu$ refers to the kinematic viscosity of the injected fluid.  To determine the parameter controlling the discrete frequency of the instability, we refer to the disturbance kinetic energy equations for this flow, which have been studied in detail by Selvam et al. \cite{Selvam2007} and Yang and Srinivasan \cite{Yang2024}. The dominant term in these equations arises from the additional stress term in the momentum equation, viz. $\frac{\partial \mu}{\partial r}\frac{\partial U_z}{\partial r}$. This term is balanced with the time derivative term on the left hand side of the axial momentum equation $\frac{\partial U_z}{\partial t}$. Further, if we make the assumption, as in \cite{Selvam2007, Yang2024} that the viscosity is an exponential function of the concentration $c$, i.e. $\nu(r) = e^{mc}$ where $m=\log(M)$, and that the thicknesses of the regions of variable viscosity $\theta _\mu$ and momentum $\theta$ scale as $\theta _\mu \sim 1/\sqrt{ReSc}$, $\theta \sim 1/\sqrt{Re}$, then the balance of terms reduces to $fD^2/\nu \sim M\log(M)Re$. However, this yields a very poor scaling. In fact a better scaling is achieved with 

\begin{equation}
\frac{fD^2}{\nu}M = A+B\cdot Re\log(M)
\end{equation}
where $A$ and $B$ are fitting constants given by $A=-5.1 \times 10^4$ and $B=23$. As seen in Fig. \ref{fig:universality}, agreement is reasonable. We reiterate that the above scaling law has no explicit dependence on the boundary layer thickness of the laminar profile entering the ambient, and the actual dependence will need further experiments in which the velocity profile shape is varied by imposing different levels of acceleration through alteration of nozzle geometry for fixed Reynolds numbers. 

\subsection{Identification of Spatial Modes}

The hot film anemometry and FFT analysis of pixel records yield similar trends for the Strouhal number as a function of $Re$ and $M$, and also show similar exponential growth of the disturbance in the downstream direction. However, they cannot answer the question of the spatial distribution of the mode, which visual observations identify as helical, axisymmetric or less clearly, transitional modes with a mix of both characteristics. This raises questions on whether these instabilities correspond to a global mode, which is usually characterized by a sharp onset as a controlling parameter (e.g. density ratio) is changed. The computations of Yang and Srinivasan \cite{Yang2024} suggest that the eigenfunctions of the absolutely unstable helical and axisymmetric modes are very similar, and that the convective growth rates are also very close. The only difference found was that the axisymmetric mode was predicted to be more unstable at nearly all combinations of $M$ and $Re$, which is not observed in experiments. However, it must be noted that the calculations predict that both modes are absolutely unstable over the parameter space covered. Therefore, it was conjectured that identification of spatial modes using a technique such as Proper Orthogonal Decomposition may elucidate the process of transition from the helical to axisymmetric modes and to detect any sharp differences in the spatio-temporal distribution that are not evident from visual observations or anemometry. 

Spectral Proper Orthogonal Decomposition (SPOD) is one of a class of methods used for identifying dominant coherent structures or spatio-temporal modes \cite{taira2017modal} that have been gaining popularity in recent years due to their ability to aid in inexpensive flow field reconstruction. A detailed explanation of the theory behind Spectral Proper Orthogonal Decomposition is available in many articles, including \cite{schmidt2020guide, towne2018spectral}. Here, we very briefly review the underlying principles that allow the use of this technique, before delving into the actual manipulation of the images acquired in experiments. SPOD is a technique to identify coherent structures (i.e. spatio-temporal modes) from data acquired through either numerical simulations or experimental measurements. Under the conditions of ergodicity and stationarity, it can be shown that the flow field can be optimally approximated by a set of basis functions that are orthogonal in the sense of an inner product that is integrated over both space and time. SPOD yields the most optimal orthogonal basis, i.e. it uses the least number of modes to capture the maximum fraction of the energy variance in the data, than any other expansion. SPOD modes are functions of both space and time, and are particularly suitable for identifying the modes associated with individual frequencies. This is because the SPOD modes are eigenvectors of a cross-spectral density tensor which is formed by the covariance of the measurements over space and time. Taking the Discrete Fourier Transform of this tensor yields equations for each individual frequency which can be solved to yield the spatio-temporal fluctuation for each frequency. 

This process is illustrated by a description of the image processing steps used in the present study. The image files were first mean-adjusted so that they contained only the fluctuation of the grayscale intensity obtained from the LIF technique. Then, the original $1024 \times 1024$ pixels images were cropped to $N_s = 484\times 132$ where the subscript s stands for `space’. An ensemble of $N_t =$ 2944 images corresponding to a time interval of 5.888 seconds were used. Each image was re-organized into a column vector of size $N_s$, and a matrix $Q$ of size $N_s\times N_t$ was constructed that has the data from all images spanning the time of observation.

Next, Welch’s method is used to calculate the Discrete Fourier Transform of the grayscale values of the fluctuation intensity. The $N_t$ temporal observations at any spatial location are split into $N_{blk}$ blocks, each containing $N_{FFT}$ images, with an overlap between blocks of $N_{FFT}/2$ images. The Discrete Fourier Transform is now applied to yield the data matrix  containing the Fourier transformed coefficients $\hat{Q}$:

\begin{equation}
\hat{Q}=\left[\begin{array}{cccc}
\mid & \mid & & \mid \\
\hat{q}^{(1)} & \hat{q}^{(2)} & \cdots & \hat{q}^{(N)} \\
\mid & \mid & & \mid
\end{array}\right]
\end{equation}

The Cross-Spectral Density tensor is given by forming the covariance matrix of $\hat{Q}$ :
\begin{equation}
    \hat{C} =\frac{1}{N_t-1}\hat{Q}^H\hat{Q}
\end{equation}

where the $\hat{Q}^H$ is the Hermitian of $\hat{Q}$. 
This equation can be written for each of the Fourier-transformed frequencies, and the corresponding spatio-temporal mode can be found by solving the singular value problem 
\begin{equation}
\hat{Q}^H\hat{Q}\hat{\Psi} = \hat{\Psi} \hat{\Lambda} 
\end{equation}

\begin{figure}[ht]
    \centering
    \begin{subfigure}{0.32\textwidth}
        \includegraphics[width=\textwidth]{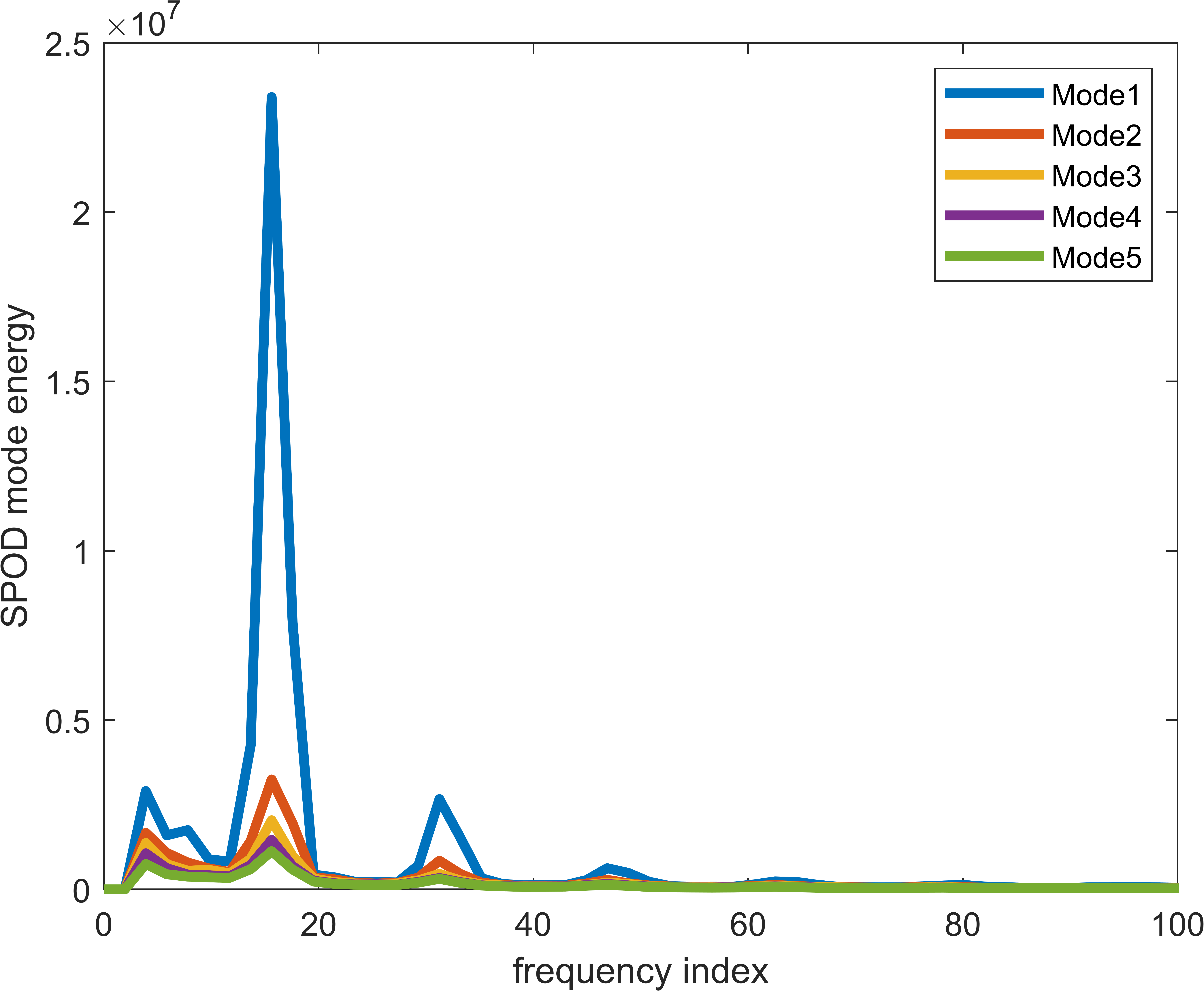}
    \caption{}
    \end{subfigure}
    \begin{subfigure}{0.32\textwidth}
        \includegraphics[width=\textwidth]{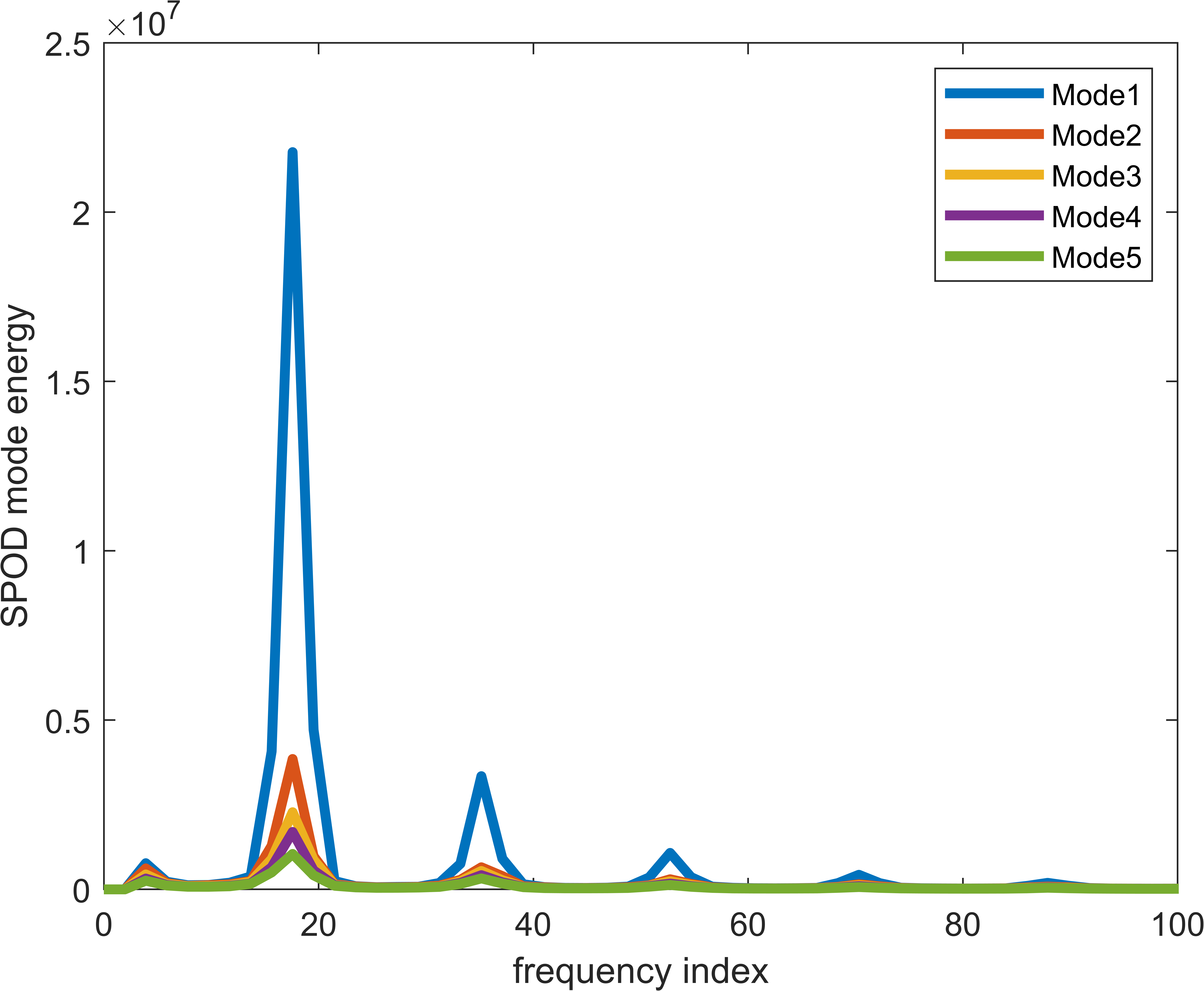}
    \caption{}
    \end{subfigure}
        \begin{subfigure}{0.32\textwidth}
        \includegraphics[width=\textwidth]{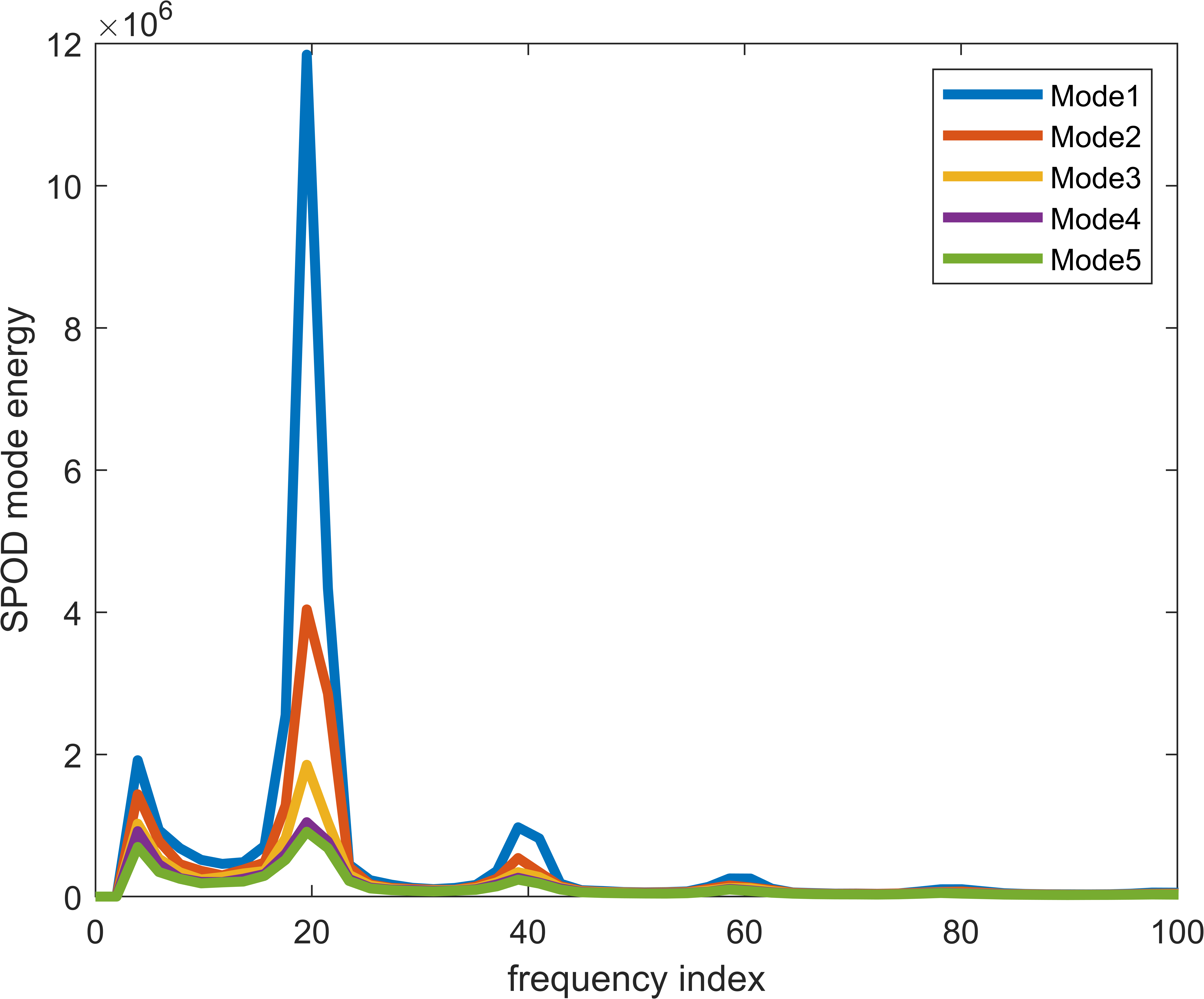}
    \caption{}
    \end{subfigure}
    \begin{subfigure}{0.32\textwidth}
        \includegraphics[width=\textwidth]{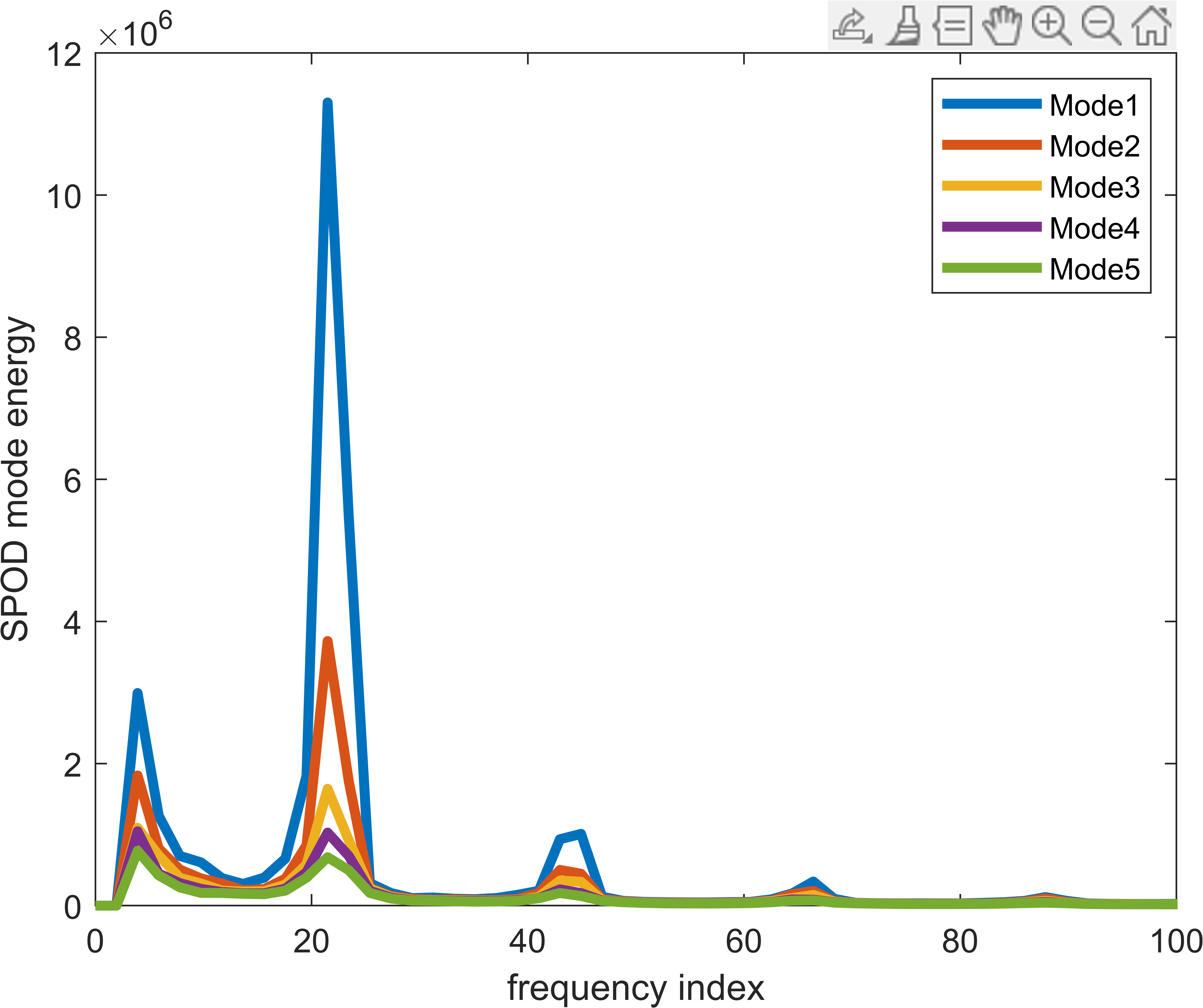}
    \caption{}
    \end{subfigure}
    \begin{subfigure}{0.32\textwidth}
        \includegraphics[width=\textwidth]{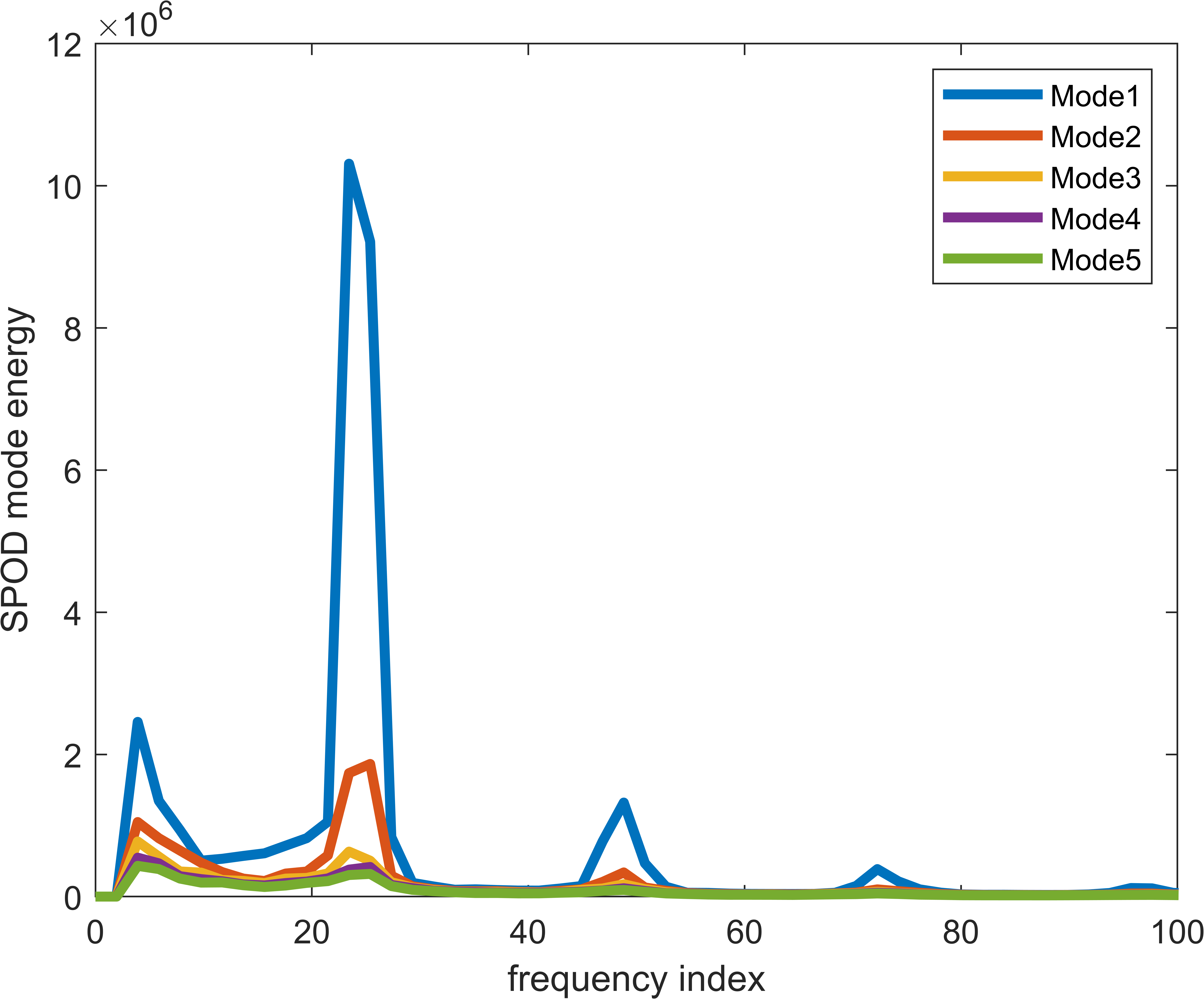}
    \caption{}
    \end{subfigure}
    \begin{subfigure}{0.32\textwidth}
        \includegraphics[width=\textwidth]{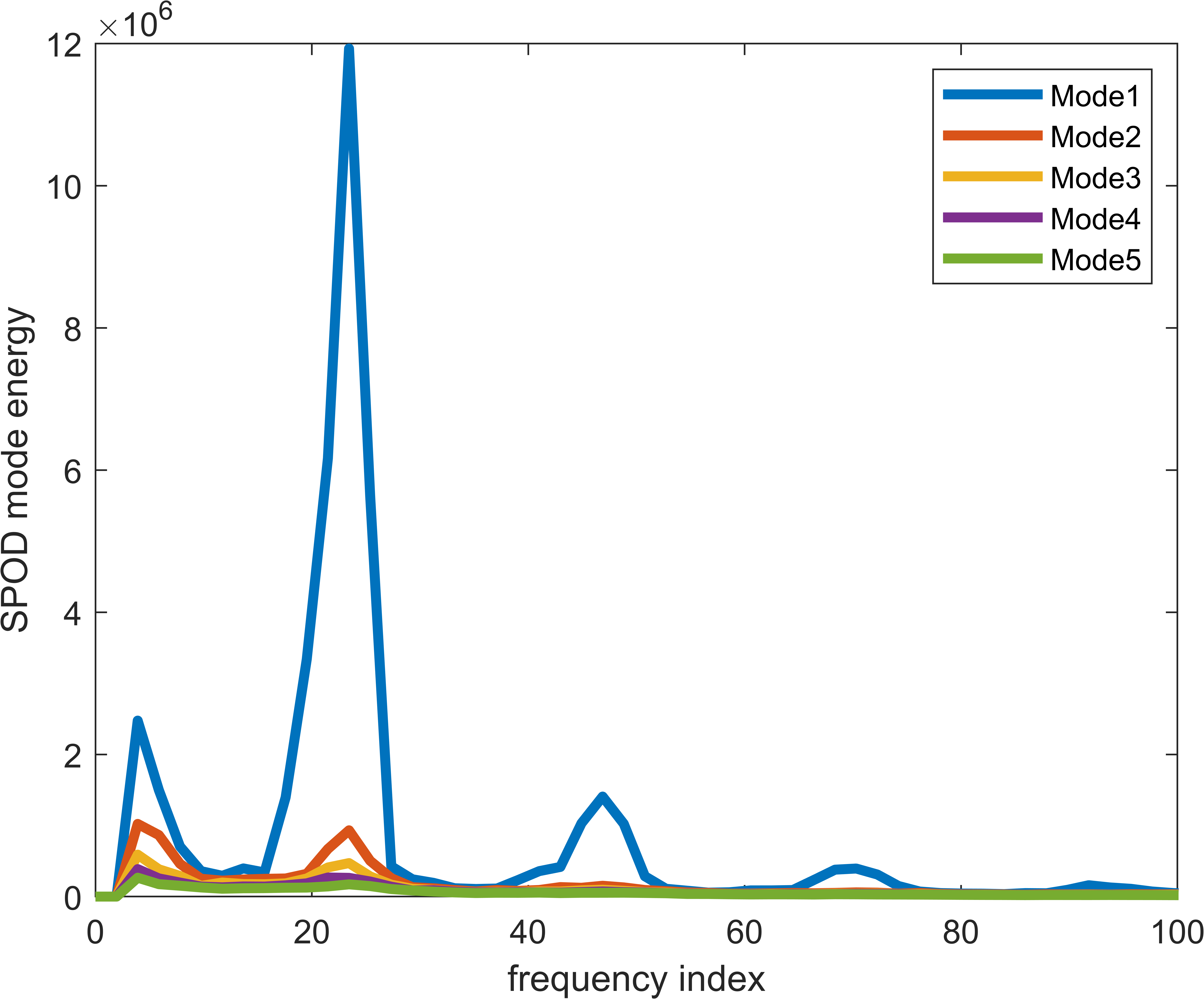}
    \caption{}
    \end{subfigure}
    \caption{The frequency spectrum of the first 6 SPOD modes with maximum energy content for trials at $Re=1600$: (a) $M=37.2$, (b) $M=33.4$, (c) $M=28.9$, (d) $M=24.7$, (e) $M=20.9$, and (f) $M=17.3$.}
    \label{fig:SPOD_spectrum}
\end{figure}

This yields the singular values $\hat{\Lambda}$ and the right singular vectors $\hat{\Psi}$ while the Fourier-transformed SPOD modes are given by $\hat{\Phi} =\hat{Q}\hat{\Psi}$. These can be inverse-transformed into the time domain and reshaped in space to get back the spatio-temporal mode in time and space. This yields $N_{blk}$ modes with a maximum resolvable frequency given by the Nyquist criterion corresponding to $N_{FFT}/2$. In the present case, following the standard recommendations, values of $N_{FFT}=128$ were used which led to $N_{blk}=23$.

\begin{figure}
    \centering
    \begin{subfigure}{0.16\textwidth}
\includegraphics[width=\textwidth]{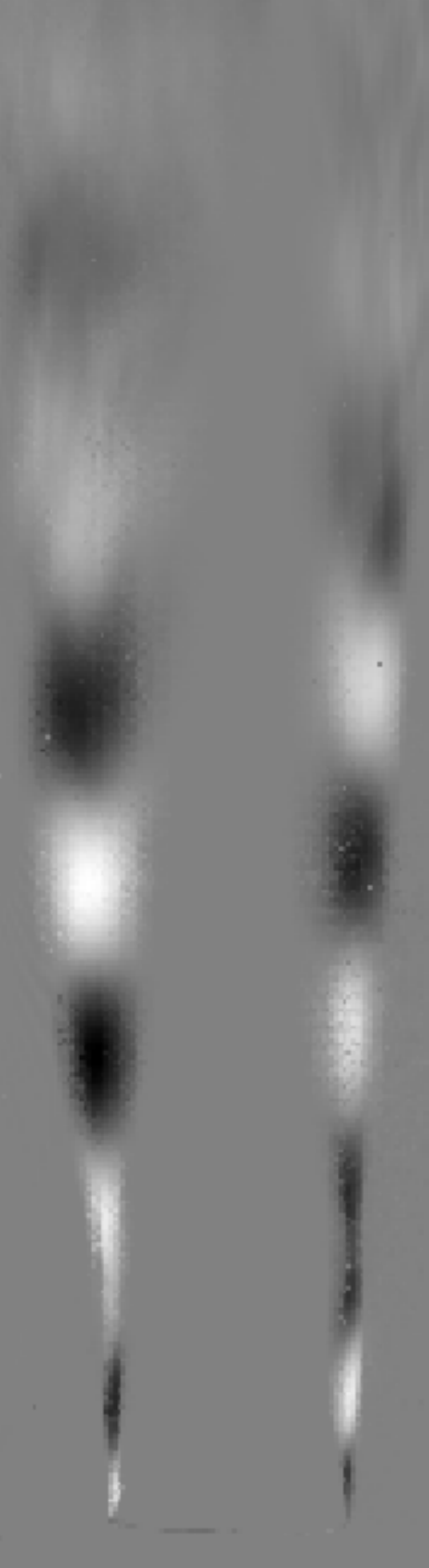}
\caption{}
\end{subfigure}
\begin{subfigure}{0.16\textwidth}
\includegraphics[width=\textwidth]{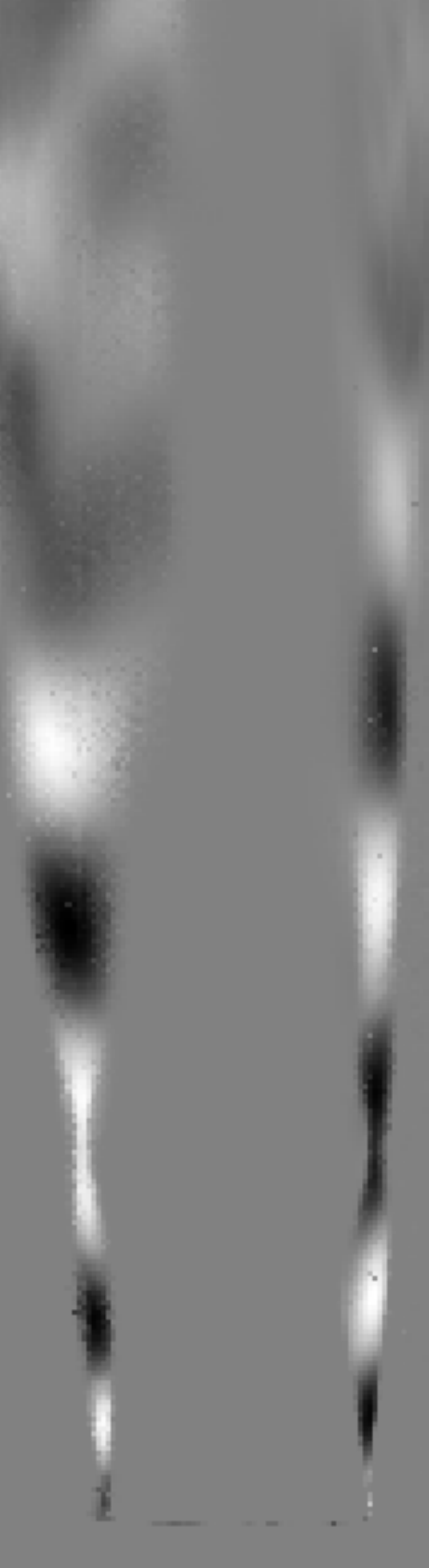}
\caption{}
\end{subfigure}
\begin{subfigure}{0.16\textwidth}
\includegraphics[width=\textwidth]{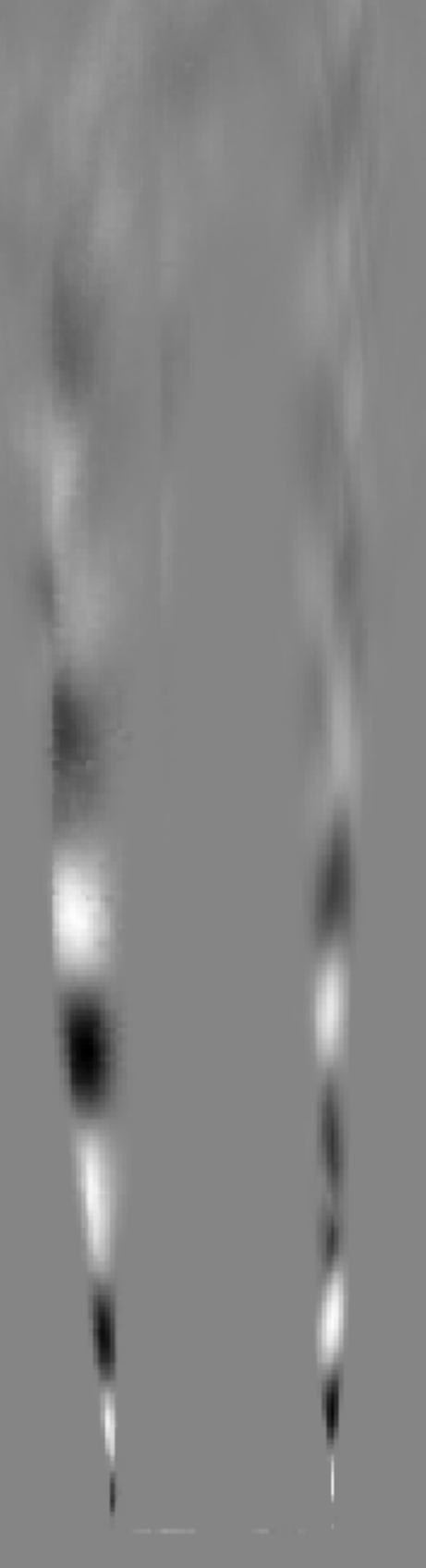}
\caption{}
\end{subfigure}
\begin{subfigure}{0.16\textwidth}
\includegraphics[width=\textwidth]{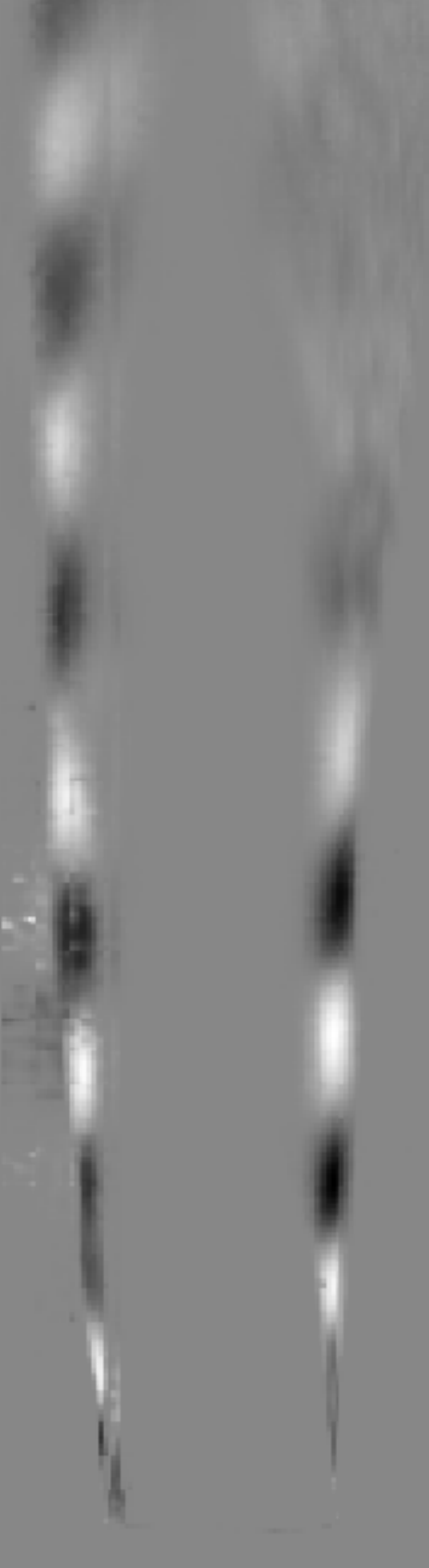}
\caption{}
\end{subfigure}
\begin{subfigure}{0.16\textwidth}
\includegraphics[width=\textwidth]{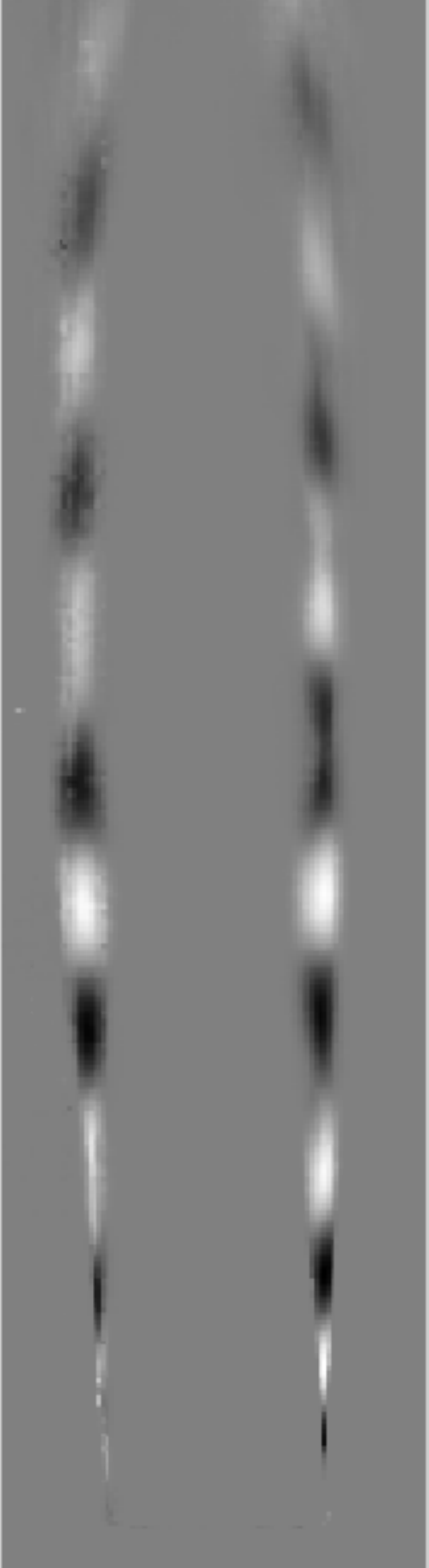}
 \caption{}
 \end{subfigure}
\begin{subfigure}{0.16\textwidth}
\includegraphics[width=\textwidth]{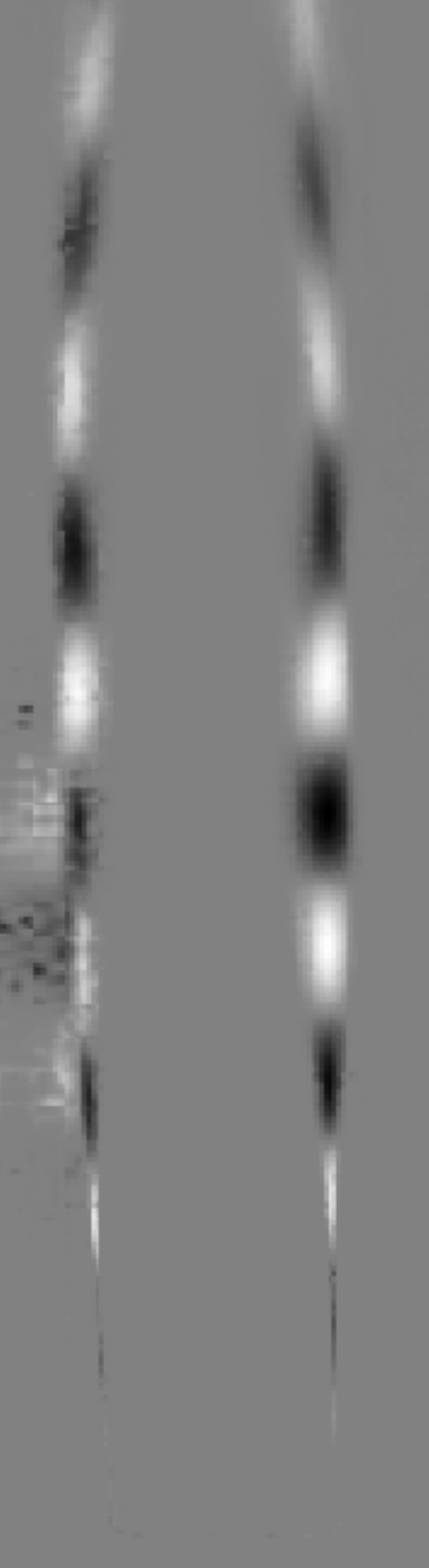}
\caption{}
\end{subfigure}
\caption{The dominant spatio-temporal mode as a function of viscosity ratio $M$ at $Re=1600$ for (a) $M=37.2$, (b) $M=33.4$, (c) $M=28.9$, (d) $M=24.7$, (e) $M=20.9$, and (f) $M=17.3$.}
\label{fig:spatialmode_Re1600}
\end{figure}

SPOD modes capture both the spatial and temporal variation, and hence videos are generated that show their behavior \cite{suppl_material}. Fig. \ref{fig:SPOD_spectrum} shows the power spectral density of the five SPOD modes with the most energy for six trials at $Re=1600$. The various subfigures (a)-(f) correspond to different trials in which $M$ is varied. In each of these spectra, it is clear that there is a dominant frequency, with little energy at other frequencies, and further, that at this dominant frequency, there is essentially one dominant spatio-temporal mode which contains most of the energy. Table 1 examines the values of these frequency peaks and compares them to the values found by performing an FFT on grayscale intensity records at single-pixel locations in the shear layer over the entire sequence of images, described in the previous subsection A (Image Analysis). It can be seen that there is a very close match for all $M$, which is not entirely surprising since both observations refer to a Fourier transform. 

\begin{table}
\centering
\begin{tabular}{lcccccc}\hline
$M$ &37.2&33.4&28.9&24.7 &20.9&17.3\\\hline
Single Pixel Analysis &15.67&17.67 &20.11&22.17&22.97& 24.17 \\\hline
SPOD &15.62 & 17.58&19.53& 22.0&23.0 &23.43\\\hline
\end{tabular}
\caption{Comparison of dominant frequencies obtained from single-pixel interrogation of grayscale fluctuations in the shear layer with the frequencies detected by SPOD analysis for trials at $Re=1600$.}
\end{table}

While the full understanding of the spatio-temporal modes at each frequency can be gleaned only from videos, still images may yet shed light on the nature of the modes and whether they are distinctively axisymmetric, helical or a combination of both. Fig. \ref{fig:spatialmode_Re1600} presents still images from the videos of the dominant spatio-temporal mode for each dominant frequency.  While the jet core and the ambient have no detail, strong patterns are evident in the shear layer. In particular, the regions of concavity/convexity of the jet boundary show up as light/dark regions. As $M$ increases from right to left, a careful examination of the images suggests that for low values of $M$, the light/dark regions at diametrically opposite locations occur at the same axial distance from the jet exit plane, suggesting an axisymmetric mode. For the highest value of $M$, these regions appear out of phase, with the light region starting at a downstream location where the dark region is present on the diametrically opposite side, suggesting a mode that is helical. For intermediate values of $M$, the light regions on one side overlap with the dark regions on the opposite side, indicating a phase shift that is not quite $180^{o}$ but still attests to the presence of a helical mode. This would suggest that the transition zone identified in Fig. \ref{fig:M_Re_plane} can be construed as a zone where the helical mode is still present, and the lower dotted line indicating the lower boundary of this zone is likely the boundary between the helical and axisymmetric modes.

\section{Summary and Discussion}

We have presented experiments characterizing the flow behavior for the specific configuration of a low-viscosity jet with a laminar boundary layer emerging into an ambient medium of relatively higher viscosity. The fluids, while perfectly miscible, have a low binary diffusion coefficient, with the result that the diffusion region is expected to be extremely thin. The flow visualization results clearly indicate the onset of a helical mode at or close to the nozzle exit plane, when the viscosity ratio $M$ is increased substantially beyond unity, with enhanced mixing relative to the $M=1$ case. For that baseline case, it is apparent that the jet can stay coherent and nearly parallel for a significant distance downstream indicating a slowly evolving spatial instability that saturates in the far-field. For large $M$, hot film anemometry  confirms the presence of a self-sustained oscillation with a discrete peak in the spectrum, with the disturbance energy rapidly growing and saturating within the first few diameters downstream. Further, we have characterized this dominant frequency as a function of $M$ and $Re$ over a limited range, showing that this frequency is an increasing function of $Re$ and a decreasing function of $M$. However, over the range of $M$ studied, the anemometry does not display any sharp jumps in the frequency, leaving open the question of the convective/absolute nature of the instability. To resolve issues with anemometry in conducting fluids, a larger range of Reynolds numbers were examined using image analysis. With this technique,the initial linear growth, followed by an exponential rise in the amplitude of oscillations is more clearly evident. We are able to confirm the frequency trends observed with anemometry,  as well as the spatial invariance of this frequency in the first few diameters downstream of the exit plane. 

Turning to the convective/absolute nature of the instability, at first glance, one cannot rule out the possibility that these observations reflect a rapidly growing convective instability that happens to be clearly visible against a low-noise background established by pump-less flow. However, there are several features of the flow, as discussed above, that suggest otherwise. It is generally accepted that for an instability to be considered a global mode that arises from a linear, absolutely unstable mechanism (as understood in the context of spatio-temporal linear stability analysis), it should display certain hallmarks. These are: the presence of a single self-sustained oscillation that can be detected everywhere in the domain; the sharp onset of a regime of enhanced mixing, as determined by the appropriate control parameters; and the insensitivity to external forcing. The first criterion is found to be amply satisfied in the present experiment. The image analysis using SPOD helps reduce the transition zone estimated from visual observations to a much narrower window corresponding to the step changes in viscosity between trials, and suggests a sharper onset than initially apparent. The transition boundary between helical and axisymmetric modes in the $M$-$Re$ plane also  corresponds qualitatively to the absolute/convective transition boundary identified through linear stability analysis (this comparison is shown in Fig. 24 of the companion paper \cite{Yang2024}). Further, the predictions of spatio-tempoal linear theory suggest that the helical and axisymmetric modes have very similar excitation frequencies, suggesting that sharp breaks in frequency may not be evident. Nevertheless, it is of interest to note that the trend of frequency with $M$ is reasonably well-captured. We also note that in line with the theoretical understanding that  that the development length for the global mode is, in principle, infinite at the transition boundary and onset of the global mode (in terms of the controlling parameters)  \citep{Couairon1997} but shortens away from the transition boundary as one moves into the regime of absolute instability, the disappearance of a region of parallel flow for large $M$, compared to $M=1$ is an indicator of the instability being controlled by inlet conditions. 

The question of sensitivity (or lack thereof) to external noise remains to be explored. Since the flow is gravity-fed, pump-driven oscillations such as those used by d'Olce et al. \cite{dOlce2009} are not considered, and current work focuses on applying vibrations to the diffuser section upstream of the nozzle. A strong response by the system, in terms of enhanced amplitude of oscillations, to forcing frequencies near the natural frequency of the instability would provide further strong support for the idea that the observed helical mode corresponds to the absolutely unstable mode found in the companion computational paper by Yang et al. \cite{Yang2024}. However, it should be noted that this may not necessarily constitute clinching evidence; indeed, Hallberg and Strykowski \cite{Hallberg2008} have shown that the global mode is a low-density jet can be overwhelmed by sufficiently strong forcing. Lastly,  it should be noted that global mode characteristics are closely linked to inlet profiles, and therefore the influence of the boundary layer thickness needs to be explicitly addressed. In the present study, this thickness is directly linked to the Reynolds number through the nozzle contraction profile, and future studies will require disentangling the relative contributions of these two parameters to the evolution of the flow.

\textbf{Acknowledgement} This research was funded by the National Science Foundation through grant \# CBET/2023932.
\bibliographystyle{unsrt}
\bibliography{refs}

\end{document}